\documentclass[11pt]{article}

\usepackage[T1]{fontenc}
\usepackage[utf8]{inputenc}

\usepackage{lmodern}
\usepackage{xspace}
\usepackage{amsfonts,amsmath,amssymb, amsthm, mathtools}
\usepackage{dsfont} 

\usepackage{algorithm}
\usepackage[noend]{algpseudocode}

\usepackage[usenames,dvipsnames,table]{xcolor}
\usepackage{multirow}

\usepackage{tikz}
\usetikzlibrary{arrows}
\usetikzlibrary{calc,decorations.pathmorphing,patterns}

\usepackage[backref,colorlinks,citecolor=blue,bookmarks=true,linktocpage]{hyperref}
\usepackage{aliascnt}
\usepackage[numbered]{bookmark}
\usepackage[capitalise]{cleveref}
\usepackage{fullpage}

\usepackage[shortlabels]{enumitem}
  \setitemize{noitemsep,topsep=3pt,parsep=2pt,partopsep=2pt} 
  \setenumerate{itemsep=1pt,topsep=2pt,parsep=2pt,partopsep=2pt}
  \setdescription{itemsep=1pt}
  
\usepackage{framed}
\usepackage{mleftright} 
\RequirePackage{pgfplots, tikz, pgflibraryplotmarks}
\usepackage{tikz,forest}
\usetikzlibrary{arrows.meta}
\PassOptionsToPackage{hyphens}{url}
\usepackage{hyperref}
 \usepackage{utopia}
 
\makeatletter
  \theoremstyle{plain} 
  	\newtheorem{theorem}{Theorem}[section]
  	\newtheorem*{theorem*}{Theorem} 
  	\newaliascnt{coro}{theorem}
  	  \newtheorem{corollary}[coro]{Corollary}
  	\aliascntresetthe{coro}
  	\newaliascnt{lem}{theorem}
  		\newtheorem{lemma}[lem]{Lemma}
  	\aliascntresetthe{lem}
  	\newaliascnt{clm}{theorem}
  		
	\aliascntresetthe{clm}
	\newaliascnt{fact}{theorem}
 	 	
	\aliascntresetthe{fact}
  	
  \newaliascnt{prop}{theorem}
  		\newtheorem{proposition}[prop]{Proposition}
	\aliascntresetthe{prop}
	\newaliascnt{conj}{theorem}
  		
	\aliascntresetthe{conj}

  \theoremstyle{remark} 
  	\newtheorem{remark}[theorem]{Remark}

  \theoremstyle{definition} 
  	\newaliascnt{defn}{theorem}
 		 \newtheorem{definition}[defn]{Definition}
 	 \aliascntresetthe{defn}

\newenvironment{proofof}[1]{\begin{proof}[\bf Proof of {#1}]}{\end{proof}}

\providecommand{\email}[1]{\href{mailto:#1}{\nolinkurl{#1}\xspace}}

\newcommand{\eps}{\ensuremath{\varepsilon}\xspace}
 
\newcommand{\Tester}{\ensuremath{\mathcal{T}}\xspace}

\newcommand{\eqdef}{:=}
\newcommand{\accept}{\textsf{accept}\xspace}

\newcommand{\reject}{\textsf{reject}\xspace}
 
\newcommand{\distribs}[1]{\Delta\mleft(#1\mright)}

\newcommand{\bigO}[1]{{O\mleft( #1 \mright)}}

\newcommand{\bigTheta}[1]{{\Theta\mleft( #1 \mright)}}
\newcommand{\bigOmega}[1]{{\Omega\mleft( #1 \mright)}}

\newcommand{\tildeO}[1]{\tilde{O}\mleft( #1 \mright)}

\newcommand{\setOfSuchThat}[2]{ \left\{\; #1 \;\colon\; #2\; \right\} } 			
\newcommand{\indicSet}[1]{\mathds{1}_{#1}}                                              
\newcommand{\indic}[1]{\indicSet{\left\{#1\right\}}}

\newcommand{\dtv}{\operatorname{d}_{\rm TV}}

\newcommand{\totalvardistrestr}[3][]{{\dtv^{#1}\mleft({#2, #3}\mright)}}
\newcommand{\totalvardist}[2]{\totalvardistrestr[]{#1}{#2}}
\newcommand{\kldiv}[2]{{\operatorname{KL}\mleft({#1 \mid\mid #2}\mright)}}
\newcommand{\chisquarerestr}[3][]{{\operatorname{d}^{#1}_{\chi^2}\mleft({#2 \mid\mid #3}\mright)}}
\newcommand{\chisquare}[2]{\chisquarerestr[]{#1}{#2}}

\newcommand{\proba}{\Pr}
\newcommand{\probaOf}[1]{\proba\mleft[\, #1\, \mright]}
\newcommand{\probaCond}[2]{\proba\mleft[\, #1 \;\middle\vert\; #2\, \mright]}
\newcommand{\probaDistrOf}[2]{\proba_{#1}\mleft[\, #2\, \mright]}

\newcommand{\expect}[1]{\mathbb{E}\mleft[#1\mright]}

\newcommand{\shortexpect}{\mathbb{E}}

\newcommand{\bE}[2]{\shortexpect_{#1}{\mleft[#2\mright]}}
\newcommand{\bEE}[1]{\bE{}{#1}}

\newcommand{\uniform}{\ensuremath{\mathbf{u}}}
\newcommand{\uniformOn}[1]{\ensuremath{\uniform_{ #1 } }}

\newcommand{\norm}[1]{\lVert#1{\rVert}}
\newcommand{\normone}[1]{{\norm{#1}}_1}
\newcommand{\normtwo}[1]{{\norm{#1}}_2}

\newcommand{\abs}[1]{\left\lvert #1 \right\rvert}
\newcommand{\dabs}[1]{\lvert #1 \rvert}
\newcommand{\dotprod}[2]{ \left\langle #1,\xspace #2 \right\rangle }

\newcommand{\clg}[1]{\left\lceil #1 \right\rceil}
\newcommand{\flr}[1]{\left\lfloor #1 \right\rfloor}

\newcommand{\R}{\ensuremath{\mathbb{R}}\xspace}

\newcommand{\N}{\ensuremath{\mathbb{N}}\xspace}

\newcommand{\lp}[1][1]{\ell_{#1}}

\algnewcommand{\LineComment}[1]{\Statex \(\triangleright\) #1}

\newcommand{\p}{\mathbf{p}}
\newcommand{\q}{\mathbf{q}}

\newcommand{\one}{\mathbf{1}}

\newcommand{\cU}{\mathcal{U}}
\newcommand{\cX}{\mathcal{X}}
\newcommand{\cY}{\mathcal{Y}}
\newcommand{\cJ}{\mathcal{J}}
\newcommand{\cP}{\mathcal{P}}
\newcommand{\cZ}{\mathcal{Z}}
\newcommand{\cW}{\mathcal{W}}
\newcommand{\cE}{\mathcal{E}}
\newcommand{\cN}{\mathcal{N}}

\newcommand{\transp}{\intercal}
\DeclareMathOperator{\Tr}{Tr}
\newcommand{\chisquaredec}{\chi^{(2)}}

\newcommand{\ouchisquaredec}{\underline{\overline{\chi}}^{(2)}}

\newcommand{\eg}{\textit{e.g.},\xspace}
\newcommand{\ie}{\textit{i.e.},\xspace}

\newcommand{\Paren}[1]{\mleft(#1\mright)}
\newcommand{\myparagraph}[1]{\medskip\noindent\textbf{#1}}

\newcommand{\ns}{n} 
\newcommand{\ab}{k} 
\newcommand{\numbits}{\ell}
\newcommand{\pbits}{s}
\newcommand{\priv}{\varrho}
\newcommand{\dst}{\eps}
\newcommand{\mech}{W}

\newcommand{\compressedab}{L}
\newcommand{\goodset}{\mathcal{G}}

\makeatletter
  \AtBeginDocument{
  \begingroup
  \toks@={}%
  \toksdef\toks@@=2 %
  \toks@@={}%
  \long\def\@ReturnFiFi#1#2\fi\fi{\fi\fi#1}%
  \def\scan@author#1#2 \and#3\@nil{%
  \ifx\\#3\\%
    \ifcase#1 %
      \toks@={#2}%
    \else
      \ifnum#1>1 %
        \toks@=\expandafter{%
          \the\expandafter\toks@\expandafter,\expandafter\space
          \the\toks@@
        }%
      \fi
      \toks@=\expandafter{\the\toks@\space and #2}%
    \fi
    \else
      \ifcase#1 %
        \toks@={#2}%
        \@ReturnFiFi{%
          \scan@author1#3\@nil
        }%
      \else
        \ifnum#1>1 %
          \toks@=\expandafter{%
            \the\expandafter\toks@\expandafter,\expandafter\space
            \the\toks@@
          }%
      \fi
      \toks@@={#2}%
      \@ReturnFiFi{%
        \scan@author2#3\@nil
      }%
    \fi
  \fi
  }%
  \expandafter\expandafter\expandafter\scan@author
  \expandafter\expandafter\expandafter0%
  \expandafter\@author\space\and\@nil
  \edef\x{\endgroup
  \noexpand\hypersetup{pdfauthor={\the\toks@}}%
  }%
  \x
  }
\makeatother

\author{
Jayadev Acharya\thanks{Supported by NSF-CCF-1846300 (CAREER).} \\
Cornell University\\
\email{acharya@cornell.edu}
\and
Cl\'ement L. Canonne\thanks{Supported by a Motwani Fellowship.}\\
Stanford University\\
\email{ccanonne@cs.stanford.edu}
\and
Yanjun Han\\
Stanford University\\
\email{yjhan@cs.stanford.edu}
\and
Ziteng Sun$^{*}$\thanks{Supported by NSF-CCF-CRII-1657471.}\\
Cornell University\\
\email{zs335@cornell.edu}
\and
Himanshu Tyagi\thanks{Supported by a grant from Robert Bosch Center for Cyber Physical Systems (RBCCPS), Indian Institute of Science.}\\ 
Indian Institute of Science\\
\email{htyagi@iisc.ac.in}
}

\title{Domain Compression and its Application to Randomness-Optimal Distributed Goodness-of-Fit}

\begin{document}
\maketitle

\begin{abstract}
We study goodness-of-fit of discrete distributions in the distributed
setting, where samples are divided between multiple users who can only
release a limited amount of information about their samples due to
various information constraints. Recently, a subset of the authors 
 showed that having access to a
common random seed (\ie{} shared randomness) leads to a significant
reduction in the sample complexity of this problem. 
In this work, we provide a complete understanding of the interplay
between the amount of shared randomness available, the stringency of
information constraints, and the sample complexity of the testing
problem by characterizing a tight trade-off between these three
parameters. We provide a general distributed goodness-of-fit protocol
that as a function of the amount of shared randomness interpolates
smoothly between the private- and public-coin sample complexities.  
We complement our upper bound with a general framework to prove lower
bounds on the sample complexity of this testing problems under limited shared
randomness.  
Finally, we instantiate our bounds for the two archetypal information
constraints of communication and local privacy, and show that
our sample complexity bounds are optimal as a
function of all the parameters of the problem, including the amount of
shared randomness.  

A key component of our upper bounds is a new primitive of \emph{domain
compression}, a tool that allows us to map distributions to a much
smaller domain size while preserving their pairwise distances, using a
limited amount of randomness. 
 \end{abstract}

\clearpage

\section{Introduction}
A prototypical example of statistical inference is that 
of \emph{goodness-of-fit}, in which one seeks to determine whether a
set of observations fits a purported probability
distribution. Considered extensively in Statistics and, more recently,
in computer science under the name of \emph{identity testing}, the
goodness-of-fit question for discrete probability distributions is by
now well-understood.    

Most of the recent work has focused on the sample complexity of the 
problem (\ie{} the minimum number of observations required to solve
the task), and sought to obtain sample-optimal, time-efficient
algorithms 
(see, \eg{}~\cite{BFRSW:10,Paninski:08,ADK:15,VV:17,DGPP:18}). In
many emerging settings, however, time or even sample considerations
may not be the main bottleneck. Instead, samples may only be partially
accessible, or their availability may be subjected to strict information
constraints. These constraints may be imposed in form of the number of
bits allowed to describe each sample (communication constraints) or
privacy constraints for each sample.  

In this context, a recent line of
work~\cite{ACT:18:IT1,ACT:19:IT2} has provided sample-optimal
algorithms under such information constraints. An important aspect
revealed by this line of work is that shared randomness is very
helpful for such problems -- public-coin protocols have much lower
sample complexity than private-coin protocols. 
However, shared randomness used by 
the distributed protocols may itself be an expensive
commodity in practice. With an eye towards practical algorithms for
deployment of these distributed statistical inference algorithms, we
consider the question of randomness-efficient distributed inference
algorithms.

Specifically, 
we consider \emph{public randomness} as a resource. In our setting,
$\ns$ users get independent samples from an unknown $\ab$-ary
distribution, and each can send a message to a central server in a
one-way, non-interactive fashion. Those messages, however, have to
comply with a prespecified \emph{local information constraint}, such
as communication (each message can be at most $\numbits$ bits long) or
local privacy (loosely speaking, messages must not divulge too much
about the user's observation.)
The server uses the $\ns$ messages to perform the goodness-of-fit test for the unknown
distribution.  

Prior work considered two natural classes of protocols: \emph{private-coin}, where users and server are randomized independently; and \emph{public-coin}, where all parties share ahead of time a common random seed that they can leverage to coordinate their messages. 
Alternatively, one may view shared randomness as the communication sent over the ``downlink'' channel by the server to the users.
In this paper, we significantly generalize prior results, by establishing a tight tradeoff between the number of users $\ns$ and the number of shared random bits $\pbits$ required for performing inference under local information constraints.

A key component of our distributed protocols is \emph{domain compression}, a new primitive we introduce.
Roughly speaking, domain compression allows one to (randomly) map a large domain $[\ab]$  to a much smaller domain of size $\compressedab\ll \ab$, while ensuring that pairwise distances between probability distributions on $[\ab]$ are (roughly) preserved when looking at their induced distributions on $[\compressedab]$. 
This notion can then be leveraged to obtain testing protocols from ``good'' domain compression mappings which use few bits of randomness. 

We proceed to describe our results in the next section, before
giving an overview of our techniques in the subsequent section. To put
our results in context, we then provide a brief overview of prior and
related work. 
   \subsection{Our Results}
  We first provide an informal overview of the setting and our results. 
We consider \emph{identity testing}, a classic example of goodness-of-fit, where one is given a reference distribution $\q$ over a known domain of size $\ab$, as well as a parameter $\dst\in(0,1)$. 
Upon receiving $\ns$ i.i.d. samples $X_1, \ldots, X_\ns$ from an unknown distribution $\p$ over the same domain, one must then output \accept with high constant probability if $\p=\q$, and $\reject$ if the total variation distance between $\p$ and $\q$ is at least $\dst$.  

\begin{figure}[h]\centering
\scalebox{.75}{\begin{tikzpicture}[->,>=stealth',shorten >=1pt,auto,node distance=20mm, semithick]
  \node[circle,draw,minimum size=13mm] (A)
  {$X_1$}; \node[circle,draw,minimum size=13mm] (B) [right of=A]
  {$X_2$}; \node (C) [right of=B] {$\dots$}; \node[circle,draw,minimum
  size=13mm] (D) [right of=C] {$X_{\ns-1}$}; \node[circle,draw,minimum
  size=13mm] (E) [right of=D] {$X_\ns$};
  
  \node[rectangle,draw,minimum width=13mm,minimum
  height=7mm,fill=gray!20!white] (WA) [below of=A]
  {$W_1$}; \node[rectangle,draw,minimum width=13mm,minimum
  height=7mm,fill=gray!20!white] (WB) [below of=B] {$W_2$}; \node (WC)
  [below of=C] {$\dots$}; \node[rectangle,draw,minimum
  width=13mm,minimum height=7mm,fill=gray!20!white] (WD) [below of=D]
  {$W_{\ns-1}$}; \node[rectangle,draw,minimum width=13mm,minimum
  height=7mm,fill=gray!20!white] (WE) [below of=E] {$W_\ns$};
  
  \node[draw,dashed,fit=(WA) (WB) (WC) (WD) (WE)] {};
  o \node[circle,draw,minimum size=13mm] (YA) [below of=WA]
  {$Y_1$}; \node[circle,draw,minimum size=13mm] (YB) [below of=WB]
  {$Y_2$}; \node (YC) [below of=WC]
  {$\dots$}; \node[circle,draw,minimum size=13mm] (YD) [below of=WD]
  {$Y_{\ns-1}$}; \node[circle,draw,minimum size=13mm] (YE) [below
  of=WE] {$Y_\ns$};

  \node (P) [above of=C] {$\p$}; \node[rectangle,draw, minimum
  size=10mm] (R) [below of=YC] {Server}; \node (out) [below
  of=R,node distance=13mm] {output};

  \draw[->] (P) edge[dotted] (A)(A) edge (WA)(WA) edge (YA)(YA) edge
  (R); \draw[->] (P) edge[dotted] (B)(B) edge (WB)(WB) edge (YB)(YB)
  edge (R); \draw[->] (P) edge[dotted] (D)(D) edge (WD)(WD) edge
  (YD)(YD) edge (R); \draw[->] (P) edge[dotted] (E)(E) edge (WE)(WE)
  edge (YE)(YE) edge (R); \draw[->] (R) edge (out);
\end{tikzpicture}
 }
\caption{The information-constrained distributed model. In the private-coin setting the channels $W_1,\dots,W_\ns$ are independent, while in the public-coin setting they are jointly randomized; in the $\pbits$-coin setting, they are randomized based on both a joint $U$ uniform on $\{0,1\}^\pbits$, and on $\ns$ independent r.v.'s $U_1,\dots,U_\ns$.}
\label{fig:model}
\end{figure}
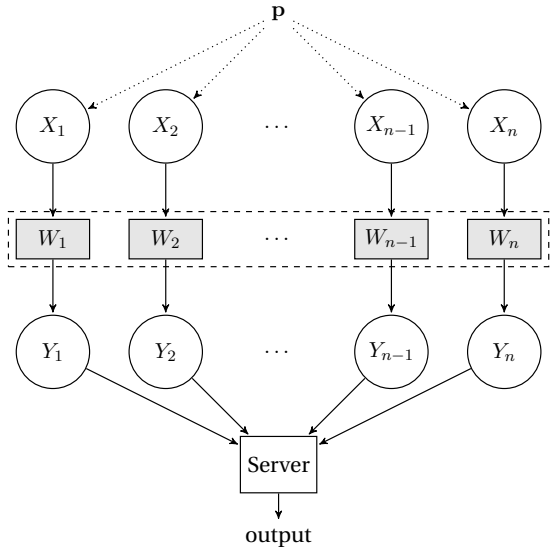

We study a distributed setting where the $X_i$'s are distributed over $\ns$ users who can only transmit a limited amount of information about their samples to a central server, which then seeks to solve the testing problem from the messages received (see~\cref{sec:preliminaries} for the detailed setup,
and~\cref{fig:model} for a pictorial description).  For simplicity, we
focus on two main applications, communication constraints and
local privacy; we point out, however, that our results are more
general, and can be leveraged to obtain both upper and lower bounds
for the more general class of information constraints described
in~\cite{ACT:18:IT1}.

\myparagraph{The communication-constrained setting.} In this setting,
each user can communicate at most $\numbits$ bits to the
server. We establish the following. 
\begin{theorem}[Informal]
  \label{theo:communication:informal} For every $\ab, \numbits \geq
1, \pbits \geq 0$, there exists a protocol for identity testing over
$[\ab]$ with $\pbits$ bits of
public randomness, $\numbits$ bits communication per user, and
\[
		  \ns
		  = \bigO{ \frac{\sqrt{\ab}}{\dst^2}\sqrt{\frac{\ab}{2^{\numbits}}\lor
		  1}\sqrt{\frac{\ab}{2^{\pbits+\numbits}}\lor 1} }\,.
\]
users. Moreover, this number of users is optimal, up to constant factors, for all values of $\ab, \pbits, \numbits$.
\end{theorem}
Note that for $\numbits \geq \log \ab$, we recover the centralized
(unconstrained) sample complexity of $O(\sqrt{\ab}/\dst^2)$; for $\pbits
= 0$ and $\pbits \geq \log \ab$, the expression matches respectively
the public- and private-coin sample complexities established in previous work.

An interesting interpretation of the sample complexity result mentioned above is that
``one bit of communication is worth two bits of public randomness.''
Equivalently, if one interprets the public randomness as an
$\pbits$ bit random seed sent over the downlink channel to the users, who then
reply with their $\numbits$-bit message, then improving the capacity
of the downlink channel is only half as effective as improving the
user-to-server channel capacities.
 
\myparagraph{The locally private setting.} In this setting, there is no
bound on the length of the message each user can send to the server,
but the randomized mechanism $\mech$ used to decide which message $y$
to send upon seeing sample $x$ has to satisfy $\priv$-local
differential privacy ($\priv$-LDP): 
\begin{equation}\label{eq:def:ldp}
    \max_{x\neq x'} \max_{y} \frac{\mech(y\mid x)}{\mech(y\mid
    x')} \leq e^{\priv}\,.
\end{equation}
(Equivalently, the probability to send any given message $y$ must stay 
roughly within a $(1\pm \priv)$ multiplicative factor, regardless of
which $x$ was observed.) We prove the following. 
\begin{theorem}[Informal]
  \label{theo:privacy:informal} For every $\ab \geq 1, \priv \in
(0,1], \pbits \geq 0$, there exists a protocol for identity testing
over $[\ab]$ under $\priv$-LDP with $\pbits$ of public randomness, and
\[
		  \ns
		  = \bigO{ \frac{\ab}{\dst^2\priv^2}\sqrt{\frac{\ab}{2^{\pbits}}\lor
		  1} }\,.
\]
users. Moreover, this number of users is optimal, up to constant factors, for all values of $\ab, \pbits$, and $\priv\in (0,1]$.
\end{theorem}
Once again, for $\pbits = 0$ and $\pbits \geq \log \ab$, this recovers 
respectively the public- and private-coin sample complexities
established in~\cite{ACFT:19,ACT:18:IT1}. In order to establish these
 upper bounds, along the way we provide  a sample-optimal
private-coin $\priv$-LDP identity testing protocol
(\cref{lemma:ldp:1bit}) which only requires one bit of communication
per user (improving in this respect on the sample-optimal protocols
of~\cite{ACFT:19}), and may be of independent interest. 

\myparagraph{General local constraints.}
Both~\cref{theo:communication:informal,theo:privacy:informal}
illustrate the versatility of our approach. To establish our
algorithmic upper bounds, we rely on a new primitive we call domain
compression (on which we elaborate in the next
subsection). Specifically, we show in~\cref{theo:upper:bounds:general}
how to combine  as a blackbox this primitive with
a \emph{private}-coin protocol for identity testing under any fixed
type of local constraint to obtain a protocol for identity testing
with $\pbits$ of public randomness, under the same local constraints.

Our proofs of optimality, similarly, are corollaries of a general
lower bound framework
(\cref{l:ic_testing_fluctuation_bound_semiprivate,theo:semimaxmin}) we
develop, and which extends that of~\cite{ACT:18:IT1} to handle limited
public randomness. We believe that both techniques~---~the domain
compression primitive, and the general lower bound formulation~--~will
find other applications in distributed statistical inference problems.
 
   \subsection{Our Techniques}
  Our proposed scheme has a modular form and, in effect, separates the
use of shared randomness from the problem of establishing an
information-constrained inference protocol. In particular, we use shared randomness only to enable {\em domain compression}.

\myparagraph{Domain compression.} 
The problem of domain compression is to convert samples from an unknown $\ab$-ary distribution $\p$ to samples from $[L]$, while preserving the total variation distances up to a factor of $\theta$. 
Our main result here is a scheme that reduces the domain-size to roughly $L\approx k\theta^2$ while preserving the total variation distance up to a factor of $\theta$. 
Furthermore, our randomized scheme does this using the optimal $2\log (1/\theta)+O(1)$ bits of randomness, which will be crucial for our applications.  
Furthermore, as we will see later, this is the best possible ``compression'' -- the lowest $L$ possible -- for a given $\theta$.

In order to come up with this optimal domain compression scheme, we establish first a one-bit $\lp[2]$ isometry for probability vectors. 
Namely, we present a random mapping which converts the domain to $\{0,1\}$ while preserving the $\lp[2]$ distances between pairs of probability vectors. 
We apply this scheme to non-overlapping parts of our $\ab$-ary probability vector to obtain the desired domain compression scheme. 
Underlying our analysis is a new anti-concentration bound for sub-Gaussian random variables, which maybe of independent interest.

\myparagraph{Domain compression to distributed testing.} With this general domain compression algorithm at our disposal, we use $\pbits$ bits of randomness to obtain a reduction of the domain size to roughly $\ab/2^\pbits$, while shrinking the statistical distances by a factor of $1/\sqrt{2^\pbits}$. Now that we have exhausted all our shared randomness in domain compression, we apply the best available private-coin protocol, but one working on domain of size $(\ab/2^\pbits)$, with new distance parameter $\dst/\sqrt{2^\pbits}$ in place of the original $\dst$.

Interestingly, when instantiating this general algorithm for specific constraints of communication, it is not always optimal to use all the randomness possible. In particular, when we have $\numbits$ bits of communication per sample available, we should compress the domain to $2^\numbits$ and use the best private-coin protocol for $\numbits$ bits of communication per sample. We formally show that {\em one bit of communication is worth two bits of shared randomness}. In particular, we should not ``waste'' any available bit of communication from the users by using too much shared randomness. 

However, this only gives us a scheme with failure probability close to $1/2$ at best. To boost the probability of error to an arbitrarily small $\delta$, the standard approach of repeating the protocol independently, unfortunately, is not an option, as we already have exhausted all available public randomness to perform the domain compression. Instead, we take recourse to a deterministic amplification technique~\cite{KPS85}, which leverages the properties of expander graphs to achieve this failure probability reduction without using any additional random bit.

\myparagraph{Optimality.} When we instantiate our general algorithm for communication and privacy constraints, we attain performance that is jointly optimal in the information constraint parameter (bits for communication and the LDP parameter for privacy), the number of samples, and the bits of shared randomness. We establish this optimality by showing chi-square fluctuation lower bounds, a technique introduced recently in~\cite{ACT:18:IT1}. This approach considers the interplay between a difficult instance of the problem and the choice of the mappings satisfying information constraints by the users. The main observation is that for public-coin protocols, the users can choose the best mapping for any given instance of the problem by coordinating using shared randomness, resulting in a minmax bottleneck. On the other hand, for private-coin protocols, for each choice of mappings, the users must handle the least favorable instance, resulting in a maxmin bottleneck. To obtain our lower bounds, we need to bridge between these two extremes and provide bounds which seamlessly switch from maxmin to minmax bounds as the number of bits of shared randomness increase. We term this significant generalization of chi-square fluctuation bounds the {\em semiminmax bound} and use it obtain tight bounds for our setting.
   \subsection{Prior and Related Work}
  Goodness-of-fit has a long and rich history in Statistics, starting with the pioneering work of Pearson~\cite{Pearson:00}. 
More recently, the composite goodness-of-fit question (where one needs to distinguish between the reference distribution, and all distributions sufficiently far in total variation from it)  has been investigated in the theoretical computer science community under the name \emph{identity testing}~\cite{GR:00,BFRSW:10}, with a focus on computational aspects and discrete distributions. 
This line of work culminated in efficient and sample-optimal testing algorithms~\cite{Paninski:08,VV:17,ADK:15,Goldreich:16,DGPP:18}; we refer the reader to the surveys~\cite{Rubinfeld:12,Canonne:15,BW:17}, as well as the recent book~\cite{Goldreich:17} (Chapter 11) for further details on identity testing, and the more general field of distribution testing. 

Recently, there has been a surge of interest in \emph{distributed} statistical inference, focusing on density or parameter estimation under communication constraints~\cite{HMOW:18,HOW:18,HMOW-ISIT:18,BHO:19} or local privacy~\cite{DuchiJW17, ErlingssonPK14,YeB17, KairouzBR16,ASZ:19,AS:19}. 
The \emph{testing} counterpart, specifically identity testing, was studied in the locally differentially private (LDP) setting by Gaboardi and Rogers~\cite{GR:18} and Sheffet~\cite{Sheffet:18}, followed by~\cite{ACFT:19}; and in the communication-constrained setting in~\cite{ACT:19:ICML,ACT:19:COLT}, as well as by (with a slightly different focus)~\cite{FMO:18}.
 The role of public randomness in distributed testing was explicitly studied in~\cite{ACT:19:ICML,ACT:19:COLT}, which showed a quantitative gap between the sample complexities of public- and private-coin protocols; those works, however, left open the fine-grained question of \emph{limited} public randomness we study here.  
 
Related to identity testing, a recent work of~\cite{DGKR:19} considers
identity testing under both memory and communication
constraints. Their setting and results, however, are incomparable to
ours, as the communication constraints they focus on are global (\ie{}
the goal is to minimize the total communication between parties), with
no hard constraint on any given user's message.

Our domain compression primitive, on the other hand, fits in the
area of \emph{dimensionality reduction}, a term encompassing
various notions whose common theme is the mapping of high-dimensional
objects into lower dimensions, while preserving (approximately) their
relevant geometric features. In our case, the objects are elements of
the $(\ab-1)$-dimensional probability simplex, and the geometric
features are the pairwise distances (mostly in $\lp[1]$
distance); this is, especially in view of our use of an $\lp[2]$
isometry to achieve this goal, reminiscent of the celebrated
Johnson-Linderstrauss (JL) lemma and its many applications~\cite{Johnson1986,IM:98}. The JL lemma, however, is for general
high-dimensional vectors, and does not necessarily map from nor into the probability
simplex. 

Closest to our primitive is the work of Kyng,
Phillips, and Venkatasubramanian~\cite{KPV:10}, which 
considers a similar question for distributions over $\R^d$ satisfying a smoothness condition.
However, their results are not applicable to our setting of finite alphabet. Furthermore, we are interested in preserving the total variation distance, and not Hellinger distance considered in~\cite{KPV:10}. Finally, our proposed algorithm is randomness efficient, which is crucial for our application. In contrast, the algorithm in~\cite{KPV:10} for domain compression requires a random mapping similar to the JL lemma construction.

\section{Notation and Preliminaries}\label{sec:preliminaries}
In what follows, we denote by $\log$ and $\ln$ the binary and natural logarithms, respectively. For an integer $\ab\geq 1$, we write $[\ab]$ for the set $\{1,\dots,\ab\}$, and $\distribs{\ab}$ for the $(\ab-1)$-dimensional probability simplex
$
    \distribs{\ab} \eqdef \{ \p\colon[\ab]\to [0,1] : \sum_{x\in[\ab]} \p(x) = 1 \}
$ (where we identify a probability distribution with its probability mass function). For $\p,\q\in\distribs{\ab}$, recall that $\totalvardist{ \p }{ \q } \eqdef \sup_{S\subseteq[\ab]} \left(\p(S)-\q(S)\right)$ is the \emph{total variation distance} between $\p$ and $\q$, which is equal to half their $\lp[1]$ distance. For our lower bounds, we shall also rely on the chi-square distance between $\p$ and $\q$, defined as $\chisquare{\p}{\q} \eqdef \sum_{x\in[\ab]} {(\p(x)-\q(x))^2}/{\q(x)}$. We indicate by $x\sim\p$ that $x$ is a sample drawn from the distribution $\p$.

 We will use standard asymptotic notations $\bigO{f}$, $\bigOmega{f}$, $\bigTheta{f}$, as well as the (relatively) standard $\tildeO{f}$, which hides polylogarithmic factors in its argument.\footnote{Specifically, $g=\tildeO{f}$ means that there exists some absolute constant $c>0$ such that $g=\bigO{f \log^c f}$.} We will, in addition, rely on the notation $a_n\lesssim b_n$  (resp. $a_n\gtrsim b_n$), to indicate there exists an absolute constant $C>0$ such that $a_n \leq C\cdot b_n$ (resp. $a_n \geq C\cdot b_n$) for all $n$, and accordingly write $a_n \asymp b_n$ when both $a_n\lesssim b_n$ and $a_n\gtrsim b_n$. Finally, for a matrix $M\in\R^{m\times n}$, we denote by $\norm{M}_F$ and $\norm{M}_\ast$ the Frobenius and nuclear norms of $M$, respectively, and by $\rho(M)$ its spectral radius.

\subsection{Setting and problem statement}
\label{ssec:setting}
In the $(\ab, \dst,\delta)$-identity testing problem, given a known reference distribution $\q\in \distribs{\ab}$, and given i.i.d. samples from $\p$, we seek to test if $\p$ equals $\q$ or if it is $\dst$-far
from $\q$ in total variation distance. Specifically, an $(\ns, \dst, \delta)$-test is given by a (randomized) mapping $\Tester\colon[\ab]^\ns\to\{0,1\}$ such that
\begin{align*}
\probaDistrOf{X^\ns\sim \p^\ns}{\Tester(X^\ns)=0} > 1-\delta &\text{ if } \p=\q,\\
\probaDistrOf{X^\ns\sim \p^\ns} {{\Tester(X^\ns)=1}} > 1-\delta &\text{ if } \totalvardist{\p}{\q}>\dst.
\end{align*}
That is, upon observing independent samples $X^\ns$, the algorithm should ``accept'' with probability at least $1-\delta$ if the samples come from the reference distribution $\q$ and ``reject'' with probability at least $1-\delta$ if they come from a distribution significantly far from $\q$. We will often fix the probability of failure $\delta$ to be a small constant, say $1/12$, and write \emph{$(\ab, \dst)$-identity testing} and \emph{$(\ns, \dst)$-test} for {$(\ab, \dst, 1/12)$-identity testing} and {$(\ns, \dst, 1/12)$-test}, respectively.\footnote{Note that the specific choice of $1/12$ is merely for convenience, and any constant less than $1/2$ would do.} The sample complexity of $(\ab, \dst)$-identity testing is the minimum $\ns$ such that we can find an $(\ns,\dst)$-test, over the worst-case reference distribution $\q$.

We work in the following distributed setting: $\ns$ users each receive an independent sample from an unknown distribution $\p\in \distribs{\ab}$, and must send a message to a central server, in the simultaneous-message-passing (SMP) setting. The local communication constraints are modeled by a family $\cW$ of ``allowed'' (randomized) channels, such that each user must select a channel $W\in\cW$ and, upon seeing their sample $x$, send the message $y=W(x)$ to the central server. Here, we focus on $\pbits$-coin SMP protocols, where the users have access to both private randomness, and a limited number of uniform public random bits. Formally, $\pbits$-coin SMP protocols are described as follows.
\begin{definition}[$\pbits$-coin SMP Protocols]\label{d:sbits-prot}
Let $U$ be an $\pbits$-bit random variable distributed uniformly over $\{0,1\}^\pbits$, independent of $(X_1, \dots, X_\ns)$; and let $U_1, \dots, U_\ns$ denote independent random variables, which are independent jointly of $(X_1, \dots, X_\ns)$ and $U$. In an \emph{$\pbits$-coin} SMP protocol, all users are given access to $U$, and further user $i$ is given access to $U_i$. 
For every $i\in[\ns]$, user $i$ selects the channel $W_i\in \cW$ as a function of $U$ and $U_i$.
The central server is given access to the random variable $U$ as well and its estimator and test can depend on $U$; however, it does not have access to the realization of $(U_1,\dots, U_\ns)$.
\end{definition}
\noindent In particular, for $\pbits=0$ we recover the private-coin setting, while for $\pbits = \infty$ we obtain the public-coin setting. 
We then say an SMP protocol $\Pi$ with $\ns$ users is an \emph{$(\ab,\dst)$-identity testing $\pbits$-coin protocol using $\cW$ with $\ns$ users} (resp. \emph{public-coin}, resp. \emph{private-coin}) if it is an $\pbits$-coin SMP protocol (resp. \emph{public-coin}, resp. \emph{private-coin}) using channels from $\cW$ which, as a whole, constitutes an $(\ns, \dst)$-test.

\medskip
\noindent\textbf{The communication-constrained and LDP channel families.} Two specific families of constraints we will consider throughout this paper are those of communication constraints, where each user can send at most $\numbits$ bits to the server, and those of $\priv$-LDP channels, where the users' channels must satisfy the definition of local differentially privacy given in~\eqref{eq:def:ldp}. We denote those two families, respectively, by $\cW_\numbits$ and $\cW_\priv$:
\[
    \cW_\numbits \eqdef \{ W\colon [\ab] \to \{0,1\}^\numbits\}, \qquad \cW_\priv \eqdef \{ W\colon [\ab] \to \{0,1\}^\ast : W \text{ satisfies }~\eqref{eq:def:ldp}\}\,.
\]

\medskip
\noindent\textbf{A useful simplification.} Throughout the paper, we will assume that the domain size $\ab$ is a power of two. This can be done without loss of generality and does not restrict the scope of our results; we establish this reduction formally in~\cref{app:useful:lemmas}. 
 
\section{Domain Compression from Shared Randomness}\label{sec:domain:compression}
We now introduce our main algorithmic tool -- a new primitive called
{\em domain compression}.  
We believe that the application of domain compression will go beyond this work. 
At a high-level, the domain compression problem requires us to convert statistical inference problems over large domain size to those over a small domain size. 
This problem is an instance of {\em universal compression}, since it is clear that we cannot assume the knowledge of the generating distribution of the samples. 
We present a simple formulation which can have applications for a variety of statistical tasks. 
Specifically, we require that pairwise distances be preserved between the distributions induced over the smaller domain. 
For our work, we only formulate a specific instance of the problem; it is easy to formulate a more general version which will have applications beyond the identity-testing problem that we consider, \eg to continuous distributions or other distance measures.

For a mapping $f\colon[\ab]\to [L]$ and $\p\in\distribs{\ab}$, denote by $\p^{f}$ the distribution of $f(X)\in [L]$.
\begin{definition}[Domain compression]
For $L<\ab$, $\cU\eqdef\{0,1\}^s$, and a mapping $\Psi\colon\cU\times
[\ab]\to [L]$, denote by $\Psi_u$, $u\in \cU$, the mapping
$\Psi_u(x)=\Psi(u,x)$.  For $\theta\in (0,1)$, the mapping $\Psi$
constitutes an {\em $(L, \theta, \delta)$-domain compression mapping}
($(L, \theta, \delta)$-DCM) for $\distribs{\ab}$ if for all
$\p, \q\in \distribs{\ab}$ such that $\totalvardist{\p}{\q}\geq \dst$,
the mapping satisfies
\begin{align}
\probaOf{\totalvardist{\p^{\Psi_U}}{\q^{\Psi_U}}\geq \theta\cdot\dst}\geq
1-\delta,
\label{e:separation_l1}
\end{align}
where the randomness is over $U$ which is distributed uniformly over $\cU$.
Furthermore, we say that this domain compression mapping uses $s$ bits of randomness.
\end{definition}
In effect, we are asking that a DCM preserves separation in total
variation distance up to a loss-factor of $\theta$ while compressing
the domain-size to $L$.  For brevity, we shall say that such a DCM
{\em compresses the domain-size} to $L$ with a {\em loss-factor} of
$\theta$.

Our main result in this section, stated below, shows that we can
compress the domain-size to $\ab \theta^2$ with a loss-factor of
$\theta$. Furthermore, we can do so using $2\log(1/\theta)$ bits of
randomness.
\begin{theorem}\label{t:domain_compression}
Suppose $\ab=2^t$ for some $t\in\N$. Then, there exist positive constants
  $c, \delta_0$ and $c_0$ such that, for every $\theta\in
  (\sqrt{c/\ab}, \sqrt{c/2})$ and every $L\geq \ab \theta^2/2c$, there is an an $(L, \theta, \delta_0)$-DCM for
  $\distribs{\ab}$. Furthermore, this domain compression mapping uses
  at most $2\log(1/\theta) + c_0$ bits of randomness.
\end{theorem}
Stated differently, we have a DCM that compresses the domain-size to
$L$ with a loss-factor of $\sqrt{L/\ab}$.  In fact, this is the
minimum loss-factor we must incur to compress the domain-size to
$L$. Indeed, by choosing $L=2^\numbits$, we can use the output of an
$(L,\theta, \delta)$-DCM to enable uniformity testing using $\numbits$
bits of communication. This output will be distributed over
$[2^\numbits]$ and the induced distribution will be separated from the
uniform distribution by at least $\theta\dst$ in total variation
distance.  Thus, using \eg{} the (non-distributed) uniformity test of~\cite{Paninski:08}, we
can complete uniformity testing using
$\sqrt{2^\numbits}/(\theta^2\dst^2)$ samples.  But this must exceed
the lower bound of $\ab/(\dst^2\sqrt{2^\numbits})$ shown
in~\cite{ACT:18:IT1} for public-coin protocols. Therefore, $\theta$
must be less than $\sqrt{2^\numbits/\ab}$. We will formalize this
proof of optimality later (see~\cref{sec:lower:bounds}), when we will show that the randomness of
$2\log(1/\theta)$ bits that we use for attaining this corner-point of
$L$ versus $\theta$ tradeoff is optimal, too. Note that we can only
achieve a constant $\delta$ from our scheme, which suffices for our
purpose. A more general treatment of the domain-compression problem,
with optimal tradeoff for all range of parameters, is an intriguing
research direction.

As described, the domain compression problem requires us to preserve
distances in total variation distance, which is equivalent to the
$\lp[1]$ metric. We have setup this definition keeping in view the
application of domain compression in identity-testing. In general, we
can consider some other metrics.  For instance, in place
of~\cref{e:separation_l1} we can require
\begin{align}
\probaOf{\normtwo{\p^{\Psi_U}-\q^{\Psi_U}}\geq
  \theta\cdot\dst}\geq 1-\delta.
\label{e:separation_l2}
\end{align}
This is a stricter requirement since $\normone{x}\geq \normtwo{x}$,
and would imply~\cref{e:separation_l1}. In fact, using a random
partition of the domain into $[L]$ parts, it was shown
in~\cite[Theorem VI.2]{ACT:19:IT2} that a loss-factor of roughly
$1/\sqrt{\ab}$ can be attained for the definition of separation
in~\cref{e:separation_l2}. This in turn implies a scheme to compress
domain-size to $L$ with a loss-factor of $1/\sqrt{\ab}$, even for the
definition of separation in~\cref{e:separation_l1}.  Comparing this
with the result of~\cref{t:domain_compression}, we find that the
performance of this random partition based DCM is off by a $\sqrt{L}$
factor from the loss-factor of $\sqrt{L/\ab}$ attained by our proposed
DCM in this paper. However, there is a simple modification than can
help: Instead of applying this scheme to the entire domain, we can
divide the domain into smaller parts and ensure $\lp[2]$ separation
for each part. If we divide the domain $[\ab]$ into equal parts and
attain $\lp[2]$ separation loss-factor of $\theta$ for each part, this
implies an overall loss-factor of $\theta$ in $\lp[1]$ as well.

To enable this approach, in the result below we establish a ``one-bit
isometry'' for $\lp[2]$ distances between distributions. That is, we
show that a random mapping $\Psi$ with one-bit output exists such that
the $\lp[2]$ distance between the distribution of output is at least
a constant times the $\lp[2]$ distance between the distribution of
input. Since the output is only binary, we can express the result in
terms of difference between probabilities of sets that map to
$1$. Note that we need this isometry not only for distribution vectors
$\p$ and $\q$, but also for subvectors of distribution vectors.
\begin{theorem}[One-bit isometry]\label{t:one-bit-isometry}
There exist absolute constants $\alpha, \delta_0, c_0$ and subsets
$\{S_u\}_{u\in \cU}$ of $[2^s]$ with $|\cU|=2^{s+c_0}$ and such that for
every $\p,\q\in [0,1]^{2^s}$ we have
\begin{align}
\probaOf{|\p(S_U) - \q(S_U)|\geq \alpha\, \normtwo{\p-\q}}\geq 1-\delta_0,
  \label{e:l2-isometry}
\end{align}
where $U$ is distributed uniformly over $\cU$ and $\p(S)\eqdef\sum_{i\in
S}\p_i$.
\end{theorem}
In other words, there is a randomized $\lp[2]$ isometry for
distributions over $[2^s]$ that uses $s+c_0$ bits of randomness. The
most significant aspect of the previous result, which is the main
workhorse for this work, is that the sets $\{S_u\}_{u\in \cU}$, are fixed
and do not depend on vectors $\p$ and $\q$.

As outlined above, we want to apply our one-bit isometry to parts of
domain. But there is one difficulty still left in implementing this
idea to obtain our desired DCM: the guarantees are only for each part
and the randomness requirement to make it work for all the parts
simultaneously maybe higher. The following simple, but useful,
observation comes to the rescue.
\begin{lemma}[Additivity of tails]\label{l:additivity_tails}
Let $a_1,\dots,a_m\geq 0$, and $Y_1, \ldots, Y_m$ be
non-negative random variables such that for some $c\in(0,1)$, $\probaOf{Y_i \geq a_i} \geq c$ for
every $1\leq i\leq m$. Then,
\[
\probaOf{Y_1+\ldots+Y_m \geq
c\cdot \frac{a_1+\ldots+a_m}{2}} \geq \frac{c}{2-c}\,.
\]	
\end{lemma}
We defer the proof of this lemma to the appendix and
of~\cref{t:one-bit-isometry} to the end of this section. For now, we
complete the proof of~\cref{t:domain_compression}, our main theorem,
using these results.

\begin{proofof}{\cref{t:domain_compression}}
Consider distributions $\p$ and $\q$ from $\distribs{\ab}$. Set $s=\clg{ \log(c/\theta^2) }$; then by our assumption, $s\leq t$.  Further,
denoting $J\eqdef2^{t-s}$, for $0\leq j \leq J-1$ define the vectors
$\p^j$ and $\q^j$ in $[0,1]^{2^s}$ as $\p^j_i\eqdef\p_{ j\cdot 2^s+i}$ and
$\q^j_i\eqdef\q_{j\cdot 2^s+i}$ for all $i\in [2^s]$. We
apply~\cref{t:one-bit-isometry} to $\p^j$ and $\q^j$ to get
\[
\probaOf{|\p^j(S_U) - \q^j(S_U)|\geq \alpha\, \sqrt{\sum_{i\in
S_U}(\p^j_i-\q^j_i)^2}}\geq 1-\delta_0, \quad 0\leq j \leq J-1.
\]
which together with the Cauchy--Schwarz inequality yields
\[
\probaOf{|\p^j(S_U) - \q^j(S_U)|\geq \alpha\cdot \frac 1{\sqrt{2^s}}
\cdot\sum_{i=1}^{2^s}|\p_{j\cdot 2^s+i}-\q_{j\cdot 2^s+i}|}\geq 1-\delta_0, \quad 0\leq j \leq J-1,
\]
We apply the ``additivity of tails''
property (\cref{l:additivity_tails}) to arrive at
\begin{align}
\probaOf{\sum_{j=0}^{J-1}|\p^j(S_U) - \q^j(S_U)|\geq \frac {2\alpha}{\sqrt{2^s}}
\cdot \totalvardist{\p}{\q}}\geq \frac{1-\delta_0}{1+\delta_0}.
\label{e:DCM_prop}
\end{align}
Consider the following function $\Psi$ with range $\{0, \dots,
2J-1\}$: For every $u\in \cU=\{0,1\}^{s+c_0}$ and $i\in [\ab]$, let
\[
\Psi(u,i)\eqdef \begin{cases}
2j, \quad &i-j\cdot 2^s\in S_u,
\\
2j+1, \quad &i-j\cdot 2^s\in [2^s]\setminus S_u,
\end{cases}
\quad 0\leq j\leq J-1.
\]
Note that $\totalvardist{\p^{\Psi_u}}{\q^{\Psi_u}}$ equals
$\sum_{j=0}^{J-1}|\p^j(S_U) - \q^j(S_U)|$. Then,~\cref{e:DCM_prop}
implies that $\Psi$ constitutes a $(2J,
2\alpha/\sqrt{2^s},2\delta_0/(1+\delta_0))$-DCM. The proof is
completed by setting $\theta\eqdef\sqrt{4\alpha^2/2^s}$ and noting that
$2J=\ab\theta^2/(2\alpha^2)$.
\end{proofof}
\begin{proofof}{\cref{t:one-bit-isometry}}
Denote $x\eqdef\p-\q$ and consider a subset $S\subseteq [2^s]$. With these
notations, the event we seek to handle is $(\sum_{i\in
S}x_i)^2\geq \alpha^2 \normtwo{x}^2$.  We associate with $S$ a vector
$u\in\{0,1\}^{2^s}$ with $i$th entry given by $\indic{i\in S}$. Then,
our of interest can be expressed as
$
x^\transp\big(uu^\transp\big)x\geq \alpha^2 \normtwo{x}^2,
$
where $\transp$ denotes the transpose. Thus, we can associate a
collection of vectors $S_1, \dots, S_m$ with a collection $u_1, \dots,
u_m$. Then, our claim can be cast as the existence of $u_1, \dots, u_m$
such that
\[
\frac 1 m \sum_{j=1}^m \indic{x^\transp\big(uu^\transp\big)x< \alpha^2 \normtwo{x}^2}
\leq \delta_0.
\]
Consider the set $\cJ$ of indices $j\in[m]$ given by
$\cJ\eqdef\{j\in [m]: x^\transp\big(u_ju_j^\transp\big)x< \alpha^2 \normtwo{x}^2\}$. 
It is easy to see that by definition of $\cJ$, we have
$
x^\transp\bigg(\frac 1
{|\cJ|}\sum_{j\in \cJ}u_ju_j^\transp\bigg)x< \alpha^2 \normtwo{x}^2,
$
which further implies
\begin{align}
\lambda_{\min}\bigg(\frac 1 {|\cJ|}\sum_{j\in \cJ}u_ju_j^\transp\bigg)< \alpha^2.
\label{e:J_set_cond}
\end{align}
The main technical component of our proof is the following result.
\begin{theorem}[Spectrum of outer products]\label{t:super_sets}
For $n\in \N$, there exist constants $c_0\in \N$, $c_1,c_2\in
(0,1)$ and vectors $u_1, \dots, u_m\in \{0,1\}^n$ with $m=2^{c_0}n$ such
that for every $\cJ \subseteq [m]$ with $|\cJ|\geq (1-c_1)m$ we must
have
\[
\lambda_{\min}\bigg(\frac 1 {|\cJ|}\sum_{j\in \cJ}u_ju_j^\transp\bigg)\geq c_2. 
\]
\end{theorem}
Specifically, we show that random binary vectors $V_1, \dots, V_m$ will
do the job.  The proof is quite technical and requires a careful analysis of
the spectrum of the random matrix $\sum_{j=1}^m V_j V_j^\transp$. In
particular, effort is required to handle entries of $V_j$ with
nonzero mean; we provide the complete proof in~\cref{app:codebook-construction}.

 We use vectors of~\cref{t:super_sets}, which implies that for vectors
$u_1, \dots, u_m$ of~\cref{t:super_sets} inequality~\eqref{e:J_set_cond}
can hold only for $|\cJ|<(1-c_1)m$. Therefore, 
\[
\frac 1 m \sum_{j=1}^m \indic{x^\transp\big(u_ju_j^\transp\big)x< c_2 \normtwo{x}^2}
\leq c_1,
\]
whereby the claim follows for sets $S_i$, $i\in [m]$, given by $S_i={\tt supp}(u_i)$ 
with $\delta_0\eqdef c_1$ and $c_2\eqdef\alpha^2$.
\end{proofof}
 
\section{Applications: Distributed Testing via Domain Compression}\label{sec:upper:bounds}
In this section, we show how the notion of domain compression
developed in~\cref{sec:domain:compression} yields distributed
protocols for identity testing under local information
constraints. Specifically, we show in~\cref{ssec:upper:bounds:general}
how to combine any private-coin identity testing protocol using $\cW$ with an
$\pbits$-coin domain compression scheme to obtain an $\pbits$-coin
identity tester using $\cW$.
Then,
in~\cref{ssec:upper:bounds:communication,ssec:upper:bounds:privacy}, we instantiate this general
algorithm with 
$\cW=\cW_\numbits$ and $\cW=\cW_\priv$ to obtain $\pbits$-coin
identity testing protocols under communication and local privacy
constraints, respectively.    

\subsection{The General Algorithm}\label{ssec:upper:bounds:general}
We establish the following result characterizing the performance of
our general algorithm.
\begin{theorem}
  \label{theo:upper:bounds:general} Let $\cW\subseteq \{ \mech \colon
  [\ab]\to\cY \}$ be a family of channels. Suppose there exists a
  $(\ab,\dst)$-identity testing private-coin protocol using $\cW$ with
  $\ns(\ab,\dst)$ players. Then, for every $s_0 < \pbits \leq \log\ab-c_0$,
  there exists a $(\ab,\dst)$-identity testing $\pbits$-public-coin
  protocol using $\cW$ with
  $C\cdot \ns(c\ab/2^\pbits,c'\dst/2^{\pbits/2})$ players, where
  $s_0, c_0,c,c'>0$ are absolute constants and $C>0$ is a constant
  depending on the desired probability of error.
\end{theorem}
      A few remarks are in order. First, we may view (and we will
illustrate this in the following sections) this statement as
saying that 
``an optimal private-coin testing protocol under local constraints
yields, \emph{as a blackbox}, an optimal $\pbits$-coin testing
protocol under the same local constraints, using domain compression.'' Second, in some cases
(such as~\cref{ssec:upper:bounds:communication}), it is 
 beneficial to use this blackbox method, 
with a number of public coins
$\pbits$ strictly smaller than the number of available public coins.
Namely, we do better by ignoring some of the shared randomness resource.
This is seemingly paradoxical, but the following heuristic may help
resolve this conundrum: reducing the domain ``too much'' may prevent the
private-coin tester from using fully what the local constraints
allow. Concretely, in the case of communication constraints where
each player can send $\numbits$ bits, reducing the domain size below
$2^\numbits$ means that some bits of communication cannot be
utilized. Third, and foremost, this theorem hints at the versatility
of our notion of domain compression and the simplicity of its use:
(i)~use public coins to reduce the domain while preserving the
pairwise distances; (ii)~run a private-coin protocol on the induced
distributions, on the smaller domain. 

\myparagraph{Overview of the proof.}
 Before delving into the details of the proof, we provide an outline of
 the argument. Suppose we have an identity testing private-coin
 protocol $\Pi$ using $\cW$. Given $\pbits$ of public randomness, we 
 use the domain compression protocol from the 
 previous section to reduce the domain size from $\ab$ to
 $\compressedab \approx \ab/2^\pbits$, while shrinking the total
 variation distances by a factor $\theta \approx
 1/\sqrt{2^\pbits}$. This entirely uses the $\pbits$ bits of public
 randomness, after which it suffices to use the private-coin $\Pi$ to
 test identity of the induced distribution
 $\p'\in\distribs{\compressedab}$ to the induced reference
 distribution $\q'\in\distribs{\compressedab}$ with distance parameter
 $\theta\cdot \dst \approx \dst/\sqrt{2^\pbits}$. Note that $\q'$ is
 known by all parties, as it is solely a function of $\q$ and the
 public randomness; and the players, after the domain compression, hold
 i.i.d. samples from $\p'$. Since the only communication between the
 parties occur when running the protocol $\Pi$ (which by assumption
 uses channels from $\cW$), the resulting protocol satisfies the local
 constraints modeled by $\cW$.
 
 This clean approach is indeed the main element of our algorithm. The
 issue, however, is that the domain compression only guarantees
 distance preservation with some constant probability $\delta_0$.
 Therefore, when $\p$ is $\dst$-far from $\q$, the approach above can only guarantee
 correctness of the overall protocol with probability at most
 $\delta_0$. In other words, the proposed protocol has low
 {\em soundness}. When $\p=\q$, however, the domain compression obviously
 yields $\p'=\q'$ with probability one, so the {\em completeness}
 guarantee holds. A standard approach to handle this would be to amplify the
 success probability by independent parallel repetitions, costing only
 a small constant factor overhead in the number of players. However, this is not
 an option for our setting, since independent repetitions would require
 fresh public randomness, \emph{which we do not have anymore.}
 Further, dividing the public randomness in different random seeds and
 using these disjoint seeds to run this amplification-by-repetition
 idea would be suboptimal, as $d$ repetitions would result in weaker
 domain compression -- we will get domain of cardinality
 $\ab/2^{\pbits/d}$ instead of the desired 
 $\ab/2^{\pbits}$.
 
To circumvent this issue, we use a different approach, that
of \emph{deterministic amplification} introduced in~\cite{KPS85}. The idea is indeed to run the protocol 
several times, say $d$, to amplify the probability of success, but
carefully reusing the \emph{same} $\pbits$ bit public randomness $U=r$
for all the $d$ runs. Namely, we can find suitable mappings
$\pi_1,\dots,\pi_d\colon\{0,1\}^\pbits\to\{0,1\}^\pbits$ such that upon
running a protocol separately for (correlated) random seeds
$\pi_1(r), \pi_2(r),\dots, \pi_d(r)$ and aggregating the results of the
$d$ distinct runs, we can amplify the success probability from $1/3$
to $\approx 1-1/d$.  Specifically, we rely on the
deterministic amplification lemma below, which guarantees that we can
drive the error from any given constant to $\delta$ paying a
factor $\tildeO{1/\delta}$ penalty in the runtime ($i.e.$, the number of parallel runs of
the protocol, and therefore also number of players), but without using
a \emph{single} extra bit of public randomness.
\begin{lemma}[Deterministic Amplification for One-Sided Error]
    \label{theo:deterministic:expansion} For any $\pbits \in\N$ and
  $\eta,\gamma\in(0,1)$, there exist $d=d(\eta,\gamma)$ and
  (time-efficiently computable) functions
  $\pi_1,\dots,\pi_d\colon\{0,1\}^\pbits\to\{0,1\}^\pbits$ such that
  the following holds.  Suppose $\cX_0\subseteq \cX$ and
  $A\colon \cX\times\{0,1\}^\pbits\to \Omega$ and
  $\cE\subseteq \Omega$ satisfy the following: 
\begin{enumerate}[(i)] \item If
  $x\in\cX_0$,
  $\probaDistrOf{\sigma \sim\{0,1\}^\pbits}{A(x,\sigma)\in\cE} =
  1$; \hfill(Perfect completeness) 
\item If $x\notin\cX_0$,
  $\probaDistrOf{\sigma \sim\{0,1\}^\pbits}{A(x,\sigma)\notin\cE} \geq
  1-\eta$. \hfill(Low soundness) 
\end{enumerate} Then, we
  have \begin{enumerate}[(i)] \item If $x\in\cX_0$,
  $\probaDistrOf{\sigma \sim\{0,1\}^\pbits}{\forall i\in[d],\,
  A(x,\pi_i(\sigma))\in\cE} = 1$; \hfill(Perfect completeness) \item If
  $x\notin\cX_0$, $\probaDistrOf{\sigma \sim\{0,1\}^\pbits}{\exists
  i\in[d],\, A(x,\pi_i(\sigma))\notin\cE} \geq 1-\gamma$. \hfill(High
  soundness)
\end{enumerate} 
Moreover, on can take
  $d=\tildeO{ \frac{\eta}{(1-\eta)^2\gamma} }$.  \end{lemma}
\noindent For completeness, we provide a self-contained proof of this
  result in~\cref{app:deterministic:amplification}. Using the lemma
  above, we can provide a straightforward algorithm to increase soundness: given
  public randomness $r\in\{0,1\}^\pbits$, we can divide the players in
  $d$ disjoint groups for some suitable (constant) $d$. Group $i$ then
  runs the natural protocol we discussed, using
  $\pi_i(r)\in\{0,1\}^\pbits$ as its public randomness; and the
  server, upon seeing the outcomes of these $d$ not-quite-independent
  protocols, aggregates them to produce the final outcome.

\begin{remark}[Universality of our algorithm]
Our proposed algorithm is universal in that the players are not
required to know the reference distribution $\q$ (in contrast to previous work~\cite{ACT:19:IT2,ACFT:19}, which relied on a reduction to uniformity testing). The same protocol
for choosing $W$s from $\cW$ works for \emph{any} identity testing
problem: the knowledge of $\q$ is only required for the center to
complete the test.
\end{remark}

We are now in position to provide the detailed proof
of~\cref{theo:upper:bounds:general}; the pseudocode of the resulting
protocol is given in~\cref{algo:full:protocol}.
\begin{proofof}{\cref{theo:upper:bounds:general}}
  We hereafter set the constants $c_0,c,\delta_0$ to be as in the
  statement of~\cref{t:domain_compression}.  Fix a reference distribution $\q\in\distribs{\ab}$, and let
  $\textsc{PrivateIdentityTesting}_{\cW}$ be a $(\ab,\dst)$-identity
  testing private-coin protocol using $\cW$ with $\ns(\ab,\dst)$
  players; with a slight abuse of notation, we will use the same name to
  invoke it with probability of failure $\delta$ for any chosen
  $\delta\in(0,1)$,
  using
  $\ns(\ab,\dst,\delta) \eqdef \ns(\ab,\dst)\cdot 72\ln(1/\delta)$
  players.\footnote{This is possible by the standard amplification
  trick: running the protocol independently several times and taking a
  majority vote. Crucially, this uses no shared randomness to perform
  as $\textsc{PrivateIdentityTesting}_{\cW}$ is private-coin.}
  Further, denote by
  $\Psi\colon \distribs{\ab}\times\{0,1\}^\pbits\to \distribs{\compressedab}$
  the $(\compressedab,\theta,\delta_0)$-domain compression mapping
  from~\cref{t:domain_compression}, where\footnote{Recall that the conditions
  of~\cref{t:domain_compression} mandate $2- \log c^2\leq
  s-c_0\leq \log \ab - \log c^2$, for some constants $c_0$ and $c$.} $\theta\eqdef
  1/\sqrt{2^{\pbits - c_0}}$ and
  $\compressedab\eqdef \ab \theta^2/2c$.
  
  By~\cref{theo:deterministic:expansion} invoked with $\eta\eqdef
 1-\delta_0$, $\gamma=1/24$, 
there exist $d
 = \Theta(1)$ and (efficiently computable)
 $\pi_1,\dots,\pi_d\colon\{0,1\}^\pbits\to\{0,1\}^\pbits$ satisfying
 the conclusion of the lemma. We will apply it to the mapping
 $A\colon \distribs{\ab}\times\{0,1\}^\pbits\to \{0,1\}$ defined by 
\[
 A(\p,r) \eqdef \indic{ \totalvardist{\p^{\Psi_r}}{\q^{\Psi_r}} \geq \theta\cdot \totalvardist{\p}{\q}
 } 
\] 
where the event $\cE$ considered is $ \cE \eqdef \{ 1 \} $, \ie
 the event that the domain compression mapping is successful. Define
 $\delta' \eqdef \min( 1-(11/12)^{1/d}, 1/24) = \Theta(1)$.  
  
  \myparagraph{The protocol.} Partition the $\ns$ players into $d$ groups
  $\Gamma_1,\dots,\Gamma_d$ of $N\eqdef \ns/d$ players, where by our
  setting of $\ns$ we have
  $N \geq \ns(\compressedab,\theta\cdot\dst,\delta')$. Given the
  public randomness $r\in\{0,1\}^\pbits$, the $N$ players in group
  $\Gamma_i$ compute their ``new'' public randomness
  $r_i \eqdef \pi_i(r)$, and use it to run the domain compression
  $\Psi$. The $N$ players in group $i$ therefore obtain i.i.d. samples
  from a distribution $\p^{(i)}\in\distribs{\compressedab}$; moreover,
  both players and server can compute the induced reference distribution
  $\q^{(i)}$ (obtained by running the domain compression $\Psi$ on
  $\q$ and randomness $r_i$). The players from $\Gamma_i$ then run the
  protocol $\textsc{PrivateIdentityTesting}_{\cW}$ on their samples
  from $\p^{(i)}$, to test identity to $\q^{(i)}$, with parameters
  $\compressedab$, $\theta\cdot\dst$ and failure probability
  $\delta'$. This results in $d$ bits $\nu_1,\dots,\nu_d\in\{0,1\}$ at
  the server, where $\nu_i$ is $0$ if the protocol run by group $i$
  returned \accept. The server then outputs $0$ (\accept) if, and only
  if, all $\nu_i$'s are equal to $0$.
  
 \myparagraph{Correctness.} First, observe that if $\p=\q$, then with
 probability one we have that $\p^{(i)} = \q^{(i)}$ for all $i\in[d]$,
 and therefore the probability that all $d$ protocols return $0$
 ($\accept$) is at least $(1-\delta')^d \geq 11/12$. This establishes
 the completeness.
 
 Suppose now that $\totalvardist{\p}{\q} > \dst$. By definition of the
 domain compression protocol and our choice of $A,\eta$, we have that
 $\probaDistrOf{r \sim\{0,1\}^\pbits}{A(\p,r)\notin\cE} \geq \delta_0
 = 1-\eta$. \cref{theo:deterministic:expansion} then guarantees that,
 with probability at least $1-\gamma = 23/24$, there exists
 $i^\ast\in[d]$ such that $\dtv(\p^{(i^\ast)},\q^{(i^\ast)})
 > \theta\cdot \dst$. When this happens, for this $i^\ast$ the
 protocol run by the players in $\Gamma_{i^\ast}$ will output
 $\nu_{i^\ast} = 1$ (\reject) with probability at least $1-\delta'$,
 and therefore by a union bound the server outputs $1$ (\reject) with
 probability at least $1-(1/24+\delta') \geq 1-1/12$.
 
\myparagraph{Number of samples.} The analysis above requires that
 $N\geq \ns(\compressedab,\theta\cdot\dst,\delta') \asymp \ns(\compressedab,\theta\cdot\dst)\log(1/\delta')$. Recalling 
 that $\ns = d\cdot N$ and that $d$, $\delta'$ are
 constant,\footnote{Specifically, if one aims for non-constant error
 probability $\delta\in(0,1)$ instead of $1/12$, we have $d
 = \tildeO{1/\delta}$ and $\delta' = \tildeO{\delta^2}$.} we have
 $\ns \asymp \ns(\ab/(2c2^{\pbits - c_0}),\dst/\sqrt{2^{\pbits -
 c_0}})$ as claimed. 
\end{proofof}

\algblockdefx[ErrorReduction]{beginErrRed}{endErrRed}
	[1]{\textbf{Deterministic error reduction} #1}
{}
\algblockdefx[DomainCompression]{beginDomRed}{endDomRed}
	[1]{\textbf{Domain compression} #1} {}
\algblockdefx[PrivateCoinTest]{beginPrivateCoinTest}{endPrivateCoinTest}
	[1]{\textbf{Private-Coin Tester} #1}{}
	
\makeatletter
\ifthenelse{\equal{\ALG@noend}{t}}
  {\algtext*{endErrRed}} {}
\ifthenelse{\equal{\ALG@noend}{t}}
  {\algtext*{endDomRed}} {}
\ifthenelse{\equal{\ALG@noend}{t}}
  {\algtext*{endPrivateCoinTest}} {}
\makeatother
\begin{algorithm}[H]\label{a:dc}
  \begin{algorithmic}[1] \Require Parameters $\ab > 1$, $\pbits\geq 1$
    with $\ab$ a power of two; $X_1,\dots,X_\ns\in[\ab]$ distributed
    among $\ns$ players, random seed $u\in\{0,1\}^\pbits$ available to
    all players.  \Ensure All players compute values
    $\compressedab,\theta$, and obtain independent samples
    $X'_1,\dots,X'_\ns\in[\compressedab]$ \State Set
    $\sigma \gets \pbits - c_0$, $\theta\gets 1/\sqrt{2^\sigma}$, and
    $\compressedab \gets \ab \theta^2/2c$ \Comment{$c_0,c$ are as
    in~\cref{t:domain_compression}.}  \State All players compute $\Psi$,
    the $(\compressedab, \theta, \delta_0)$-DCM for $\distribs{\ab}$
    guaranteed
    by~\cref{t:domain_compression}.  \For{$j\in[\ns]$} \Comment{As
    $2^\pbits = c_0\cdot 2^{\sigma}$, the players interpret $u$ as a
    random seed for $\Psi$.}  \State Player $j$ maps their sample
    $X_j\in[\ab]$ to $ X'_j \gets \Psi_U(X_j) $ in
    $[\ab]$.  \EndFor \State \Return $\compressedab\in\N$,
    $\theta\in(0,1]$ \end{algorithmic} \caption{\label{algo:domain:reduction}Domain
    compression protocol \textsc{DomainCompression}}
\end{algorithm}
\begin{algorithm}[ht!]
  \begin{algorithmic}[1] 
\Require Parameters $\ab > 1$, $\pbits\geq 0$
    with $\ab$ a power of two; $\dst,\delta \in (0,1)$ 
\Require
    Private-coin protocol $\textsc{PrivateIdentityTesting}_{\cW}$
    using $\cW$ with $\ns(\ab,\dst,\delta)$ players 
\Ensure This is a
    $(\ab,\dst,\delta)$-identity testing protocol as long as $\ns \geq
    C_\delta \cdot \ns(\ab,\dst,\delta)$ 
\beginErrRed \State
    Apply~\cref{theo:deterministic:expansion} to $\eta\eqdef 3/4$,
    $\gamma=\delta/2$, to obtain mappings
    $\pi_1,\dots,\pi_d\colon \{0,1\}^{\pbits}\to \{0,1\}^{\pbits}$ \State
    Partition the $\ns$ players into $d$ groups
    $\Gamma_1,\dots,\Gamma_d\subseteq [\ns]$ of $N\gets \ns/d$
    players \State Set
    $\delta' \gets \min(1-(1-\delta)^{1/d}, \delta/2)$ \endErrRed \beginDomRed \LineComment{The
    constants $c_0,c$ are as
    in~\cref{t:domain_compression}.}  \If{$\pbits >
    c_0$} \Comment{Enough public coins are available.}  \State All
    players agree on a uniformly random
    $R\in\{0,1\}^{\pbits}$. \Comment{This uses $\pbits$ public
    coins.}  \For{$i\in[d]$} \Comment{players in group $i$ run the
    protocol on randomness $\pi_i(R)$} \State
    $(\compressedab,\theta)\gets \textsc{DomainCompression}(\ab,\pbits,
    (X_j)_{j\in \Gamma_i} ,\pi_i(R))$ \EndFor \Else \Comment{If too
    few public coins are available, use directly the private-coin
    protocol.}  \State $(\compressedab,\theta)\gets
    (\ab,1)$ \For{$j\in[\ns]$} player $j$ sets $ X'_j \gets X_j
    $. 
\Comment{Keep same    sample} 
\EndFor \EndIf \endDomRed \beginPrivateCoinTest 
\For{$i\in[d]$} 
\Comment{Each    group runs the private-coin identity testing protocol using
    $\cW$} 
\State Let $\q^{(i)}$ be the
    reference distribution induced by $\textsc{DomainCompression}$ run
    on $\pi_i(R)$. 
 \State
    $\nu_i \gets \textsc{PrivateIdentityTesting}_{\cW}(\q^{(i)}, \compressedab, \theta\cdot\dst, \delta',
    (X'_j)_{j\in \Gamma_i})$ \Comment{$\nu_i=0$ if the test
    accepts} \EndFor \endPrivateCoinTest \State \Return $0$ (\accept)
    if $\nu_i=0$ for all $i\in[d]$, $1$ (\reject)
    otherwise \end{algorithmic} \caption{\label{algo:full:protocol}The
    full $(\ab,\dst,\delta)$-identity testing protocol}
\end{algorithm}

In the two next subsections, we will illustrate the versatility
of~\cref{theo:upper:bounds:general} by applying it to $\numbits$-bit
local communication constraints and $\priv$-local privacy constraints,
respectively, to obtain sample-optimal protocols.

\subsection{Communication-Constrained Testing}\label{ssec:upper:bounds:communication}
In the communication-constrained setting each
player can only send $\numbits < \log\ab$ bits to the server: \ie{}
$\cW=\cW_\numbits$, where $ \cW_\numbits = \{ \mech\colon
[\ab]\to\{0,1\}^\numbits \} $. We establish the
following theorem:
\begin{theorem}
		\label{thm:identity:testing:communication:ub} For any
		integers $\numbits \geq 1, \pbits \geq 0$, there
		exists an $\numbits$-bit communication protocol with
		$\pbits$ bits of public randomness using \[ \ns
		= \bigO{ \frac{\sqrt{\ab}}{\dst^2}\sqrt{\frac{\ab}{2^{\numbits}}\lor
		1}\sqrt{\frac{\ab}{2^{\pbits+\numbits}}\lor 1} } \]
		players to perform $(\ab,\dst)$-identity testing. In
		particular, for $\numbits+\pbits \leq \log\ab$, this
		becomes
		$\bigO{ \frac{\ab}{2^{\numbits/2}\dst^2}\sqrt{\frac{\ab}{2^{\pbits+\numbits}}}
		}$.
\end{theorem}
\noindent As we shall see in~\cref{sec:lower:bounds}, this is sample-optimal.
\begin{proofof}{\cref{thm:identity:testing:communication:ub}}
We note first that for $\numbits \geq \log\ab$, the setting becomes
equivalent to the centralized setting, and the claimed expression
becomes $O(\sqrt{\ab}/\dst^2)$, the (known) tight centralized sample
complexity. Thus, it is sufficient to focus on $1\leq \numbits
< \log\ab$, which we hereafter do.  To
apply~\cref{theo:upper:bounds:general}, we utilize the {\em
simulate-and-infer} private-coin identity testing protocol
of~\cite{ACT:19:IT2}.
Specifically, we invoke the following
result from~\cite{ACT:19:IT2}, which gives a
sample-optimal private-coin identity testing protocol $\Pi_\numbits$
using $\cW_\numbits$:
\begin{theorem}[{\cite[Corollary IV.3]{ACT:19:IT2}}]
    \label{theo:private:coin:identity:testing} For any integer
  $\numbits \geq 1$, there exists a private-coin
  $(\ab,\dst,\delta)$-identity testing protocol using $\cW_\numbits$
  and \[ \ns
  = \bigO{ \frac{\ab}{2^{\numbits}\dst^2} \Paren{ \sqrt{\ab\log\frac{1}{\delta}}
  + \log\frac{1}{\delta} } } \] players. In particular, for constant
  $\delta$ this becomes $\ns(\ab,\dst)
  = \bigO{ \frac{\ab^{3/2}}{2^{\numbits}\dst^2} }$.
\end{theorem}
Armed with this protocol $\Pi_\numbits$, we proceed as follows. Set
$\bar{\pbits}\gets \min(\log (\ab)-\numbits, \pbits)$ to be the
``effective'' number of usable public coins (intuitively, if more than
$\log \ab-\numbits$ public coins are available, it is not worth using
them all, as compressing the domain below $2^\numbits$ would render
some of the $\numbits$ available bits of communication useless).
\begin{itemize}
  \item If $\bar{\pbits}\leq c_0$ (where $c_0$ is the constant from
  the statement of~\cref{t:domain_compression}), then we simply run the
  private-coin protocol $\Pi_\numbits$. This
  requires \[ \ns \geq \ns(\ab,\dst) \asymp \frac{\ab^{3/2}}{2^{\numbits}\dst^2}
  = \frac{\sqrt{\ab}}{\dst^2}\cdot \sqrt{\frac{\ab}{2^{\numbits}}}\cdot \sqrt{ \frac{\ab}{2^{\numbits}}} \asymp \frac{\sqrt{\ab}}{\dst^2}\cdot \sqrt{\frac{\ab}{2^{\numbits}}}\cdot \sqrt{ \frac{\ab}{2^{\numbits
  + \pbits}} \lor 1 }, \] since either $\pbits \leq c_0$ (in which case
  $\frac{\ab}{2^{\numbits}} \asymp \frac{\ab}{2^{\numbits+\pbits}}$)
  or $\log (\ab) - \numbits \leq c_0$ (in which case
  $\frac{\ab}{2^{\numbits+\pbits}} \leq \frac{\ab}{2^{\numbits}} \lesssim
  1$).  \item Else, we apply~\cref{theo:upper:bounds:general} with
  $\bar{\pbits}$ bits of public randomness and private-coin identity
  testing protocol $\Pi_\numbits$. This can be done as long
  as 
\begin{align*}
 \ns \geq
  C\cdot \ns(c\ab/2^{\bar{\pbits}},c'\dst/2^{\bar{\pbits}/2}) \asymp \frac{(\ab/2^{\bar{\pbits}})^{3/2}}{2^{\numbits}(\dst/2^{\bar{\pbits}/2})^2}
  = \frac{\ab^{3/2}}{2^{\numbits}\dst^2 2^{\bar{\pbits}/2}}
  &= \frac{\sqrt{\ab}}{\dst^2}\cdot \sqrt{\frac{\ab}{2^{\numbits}}}\cdot \sqrt{ \frac{\ab}{2^{\numbits
  + \bar{\pbits}}}}
\\
  &= \frac{\sqrt{\ab}}{\dst^2}\cdot \sqrt{\frac{\ab}{2^{\numbits}}}\cdot \sqrt{ \frac{\ab}{2^{\numbits
  + \pbits}} \lor 1 },
\end{align*}
where  the last identity holds since
  $\bar{\pbits} = (\log (\ab)-\numbits) \land \pbits$.
\end{itemize}
This concludes the proof.
\end{proofof}

\subsection{Locally Differentially Private Testing}\label{ssec:upper:bounds:privacy}
In this section, we consider the locally private channel family, where each player can 
only send a message that is $\priv$-LDP. That is,
recalling~\cref{eq:def:ldp}, we consider the channel family
\[
	\cW = \cW_\priv = \setOfSuchThat{W\colon [\ab] \to \cY
	}{ \forall y \in \cY, \forall x_1, x_2 \in [\ab], W(y\mid
	x_2) \le e^\priv W(y\mid x_1) }\,.
\]
We establish the following result for performance of our proposed
 general algorithm for testing under privacy constraints. It will be
 seen in the next section that, much like the
 communication-constrained setting,
 for the privacy-constrained setting as well our general algorithm is
 optimal. 
\begin{theorem}
	\label{thm:test_ldp} For any integers $\ab\geq 1, \pbits \geq
	0$, and parameter $\delta\in(0,1), \priv > 0$, there exists a
	one-bit communication $\priv$-LDP protocol with $\pbits$ bits
	of public randomness using \[ \ns
	= \bigO{ \frac{\sqrt{\ab}}{\dst^2} \frac{\sqrt{\ab}}{\priv^2} \sqrt{\frac{\ab}{2^{\pbits}}\lor
	1} } \] players to perform $(\ab,\dst, \delta)$-identity
	testing. When $\pbits >\log k$, this becomes
	$\bigO{ \frac{\ab}{\dst^2\priv^2} }$.
\end{theorem}
In~\cite{ACFT:19}, it was shown that the sample complexity for
identity testing with $\priv$-local differential privacy constraints
is $\Theta( {\ab^{3/2}}/({\dst^2 \priv^2 }) )$ using only private
randomness and $\Theta( {\ab}/({\dst^2 \priv^2}) )$ with
(unlimited) public randomness.\footnote{Although, as the authors
showed, their protocol could be made to work with $O(\log\ab)$ bits of
public randomness.}{} \cref{thm:test_ldp} matches these bounds in both
cases. Moreover, we note here that for private-coin schemes, we can
achieve the optimal sample complexity with a one-bit communication
protocol. This is in contrast with the private-coin protocols
of~\cite{ACFT:19} which require $\bigOmega{\log \ab}$ bits of
communication per player. This also shows that, unlike the communication-constrained setting, under LDP constraints there is no tradeoff between the number of available bits of communication and sample complexity.

\begin{proof}	
	We will rely on the following lemma, which improves on the
	private-coin protocol of~\cite{ACFT:19} in terms of
	communication complexity (while achieving the same sample
	complexity). The protocol is inspired by that of~\cite{AS:19}, which provides a one-bit LDP protocol for
	distribution \emph{learning}.  
\begin{lemma}
  \label{lemma:ldp:1bit}
	There exists a one-bit communication private-coin $\priv$-LDP
	protocol that uses \[ \ns
	= \bigO{ \frac{\ab^{3/2}}{\dst^2\priv^2}\log\frac{1}{\delta}} \]
	players to perform $(\ab,\dst, \delta)$-identity testing. For
	constant $\delta$ this becomes $\ns_\priv(\ab,\dst)
	= \bigO{ \frac{\ab^{3/2}}{\dst^2\priv^2}}$.
\end{lemma}
We defer the proof for this intermediate result to~\cref{ssec:protocol:ldp:faster:stronger:cleaner}, and continue the proof of the theorem assuming the statement. Let us denote by $\Pi_\priv$ the protocol
	from~\cref{lemma:ldp:1bit}; we then proceed as
	follows: \begin{itemize} \item If $s \leq c_0$ (where $c_0$ is
	the constant from the statement
	of~\cref{t:domain_compression}), then we just run the
	private-coin protocol $\Pi_\priv$.  \item Else, we
	apply~\cref{theo:upper:bounds:general} with $\bar{\pbits}
	= \min(\log \ab, s)$ bits of public randomness and
	private-coin identity testing protocol $\Pi_\priv$. This can
	be done as long as \[ \ns \geq
	C\cdot \ns_\priv(c\ab/2^{\bar{\pbits}},c'\dst/2^{\bar{\pbits}/2}) \asymp \frac{(\ab/2^{\bar{\pbits}})^{3/2}}{\priv^2
	(\dst/2^{\bar{\pbits}/2})^2} = \frac{\ab^{3/2}}{\priv^2 \dst^2
	2^{\bar{\pbits}/2}} 
	= \frac{\sqrt{\ab}}{\dst^2}\cdot \frac{\sqrt{\ab}}{\priv^2}\cdot \sqrt{ \frac{\ab}{2^{\pbits}} \lor
	1 } \] the last equality recalling that $\bar{\pbits} =
	(\log \ab) \land \pbits$.  \end{itemize}
\end{proof}

\subsubsection{Proof of~\cref{lemma:ldp:1bit}}\label{ssec:protocol:ldp:faster:stronger:cleaner}

It only remains to prove~\cref{lemma:ldp:1bit}, our intermediary result giving a communication-efficient private-coin protocol for identity testing under LDP. We emphasize that the main advantage of this protocol is that we require only one bit of communication per player as compared to $\Omega(\log \ab)$ for those of~\cite{ACFT:19}, while in terms of sample complexity both protocols are optimal.
	\begin{proofof}{\cref{lemma:ldp:1bit}}
	We use the same response scheme as
		in~\cite{AS:19}. The scheme is the following. Let
		$K \eqdef 2^{ \clg{\log_2(\ab+1)} }$, which is the
		smallest power of two larger than $\ab$. Let $H_K$ be
		the $K \times K$ Hadamard matrix. Without loss of
		generality, we assume $K$ divides $\ns$ (as otherwise
		we can ignore the last $(\ns - K \flr{\frac{\ns}{K}})$
		players). Deterministically partition divide the players
		into $K$ disjoint blocks of equal size $B_1,
		B_2, \dots, B_K$. Each player $i \in B_j$ is assigned
		the $j$th column of the Hadamard matrix. Let $C_j$ be
		the location of $+1$'s on the $j$th column; the
		channel used by player $i \in B_j$ is given
		by
		\begin{equation} \label{eqn:response}
		\mech_i(1 \mid x ) =
		  \begin{cases}
		    \frac{e^\priv}{e^\priv+1}, \text{ if } x \in C_j, \\
		    \frac{1}{e^\priv+1}, \text{ otherwise.} 
		  \end{cases}
		\end{equation}
		Then, following the same computations as
		in~\cite{AS:19}, we have that for all $j \in [K]$, 
		\begin{equation*}
		p_{C_j} \eqdef \probaCond{ Y_i = 1 }{ i \in B_j}
		= \frac{e^\priv - 1}{e^\priv +1} \p(C_j)
		+ \frac{1}{e^\priv + 1}\,.  
		\end{equation*}
		Taking one player from
		each block and viewing the resulting collection of
		messages as a length-$K$ vector, we thus get $\ns/K$
		samples from a product distribution on $\{0,1\}^K$
		with mean vector $p_C\eqdef
		(p_{C_1},\dots,p_{C_K})$.  
		From a Parseval-based
		argument analogous to~\cite{ACFT:19}, we then know
		that \[ \normtwo{p_C - q_C }^2 = \frac{K(e^\priv -1)^2
		}{4(e^\priv + 1)^2} \normtwo{\p-\q}^2\,, \] where
		$q_C \in[0,1]^K$ is the mean vector obtained as above
		when the input distribution is $\q$ instead of
		$\p$. (Note that $q_C$ can be explicitly computed
		given knowledge of $\q$.)  Therefore, when
		$\totalvardist{\p}{\q} > \dst$, $\normtwo{p_C- q_C }^2
		> \alpha\eqdef\frac{(e^\priv -1)^2 }{(e^\priv +
		1)^2} \dst^2$, while $\normtwo{p_C - q_C }^2 = 0 $
		when $\p = \q$.  Since, for product distributions over
		$\{0,1\}^K$, the problem of testing whether the mean
		vector is either (i)~a prespecified vector
		$\mu\in\R^K$ or (ii)~at $\lp[2]$ distance at least
		$\alpha$ from $\mu$ has sample complexity
		$\bigO{\sqrt{K}\log(1/\delta)/\alpha^2}$,\footnote{This
		is more or less folklore; see \eg{}~\cite[Section
		2.1]{CDKS:17}, or~\cite[Lemma 4.2]{CKMUZ:19}.} having
		$\ns/K \gtrsim \sqrt{K}\log(1/\delta)/\alpha^2$
		suffices, \ie \[ \ns = \bigO{\ab^{3/2}\frac{(e^\priv +
		1)^2 }{4 (e^\priv - 1)^2 \dst^2}\log\frac{1}{\delta} }
		= \bigO{ \frac{\ab^{3/2}}{\dst^2\priv^2}\log\frac{1}{\delta}}\,, \]
		as claimed. 		
		Finally, the fact that this protocol does, indeed, satisfy the $\priv$-LDP constraints is immediate from~\cref{eqn:response}.
		\end{proofof}

\section{Lower Bounds}\label{sec:lower:bounds}
Our lower bounds consist of the following ingredients. In~\cref{ssec:semimaxmin}, we introduce the notion of \emph{semimaxmin chi-square fluctuation} of a family of channels $\cW$, which will be central to our results. 
In~\cref{theo:semimaxmin} we provide an upper bound on the semimaxmin chi-square fluctuation as a function of $\norm{H(W)}_\ast$. We then, in the corresponding following sections, use~\cref{l:ic_testing_fluctuation_bound_semiprivate}, in conjunction with the bounds on  $\norm{H(W)}_\ast$ for communication-constrained and locally private channels, to prove our lower bounds in those two settings and establish the lower bound part of~\cref{theo:communication:informal,theo:privacy:informal}.

As we aim to prove a lower bound on the sample complexity of identity testing (for general reference distribution $\q$), it is enough to show a lower bound on its special case of \emph{uniformity} testing. This is a sensible choice, as the uniform distribution $\uniformOn{\ab}$ is the ``hardest'' instance of identity testing (see \eg{}~\cite{Paninski:08,Goldreich:16}).

\subsection{The General Formulation: Semimaxmin decoupled chi-square fluctuation}\label{ssec:semimaxmin}
We build on the notions of maxmin and minmax {\em decoupled chi-square fluctuations}, introduced
in~\cite{ACT:18:IT1} to prove lower bounds on the sample
complexity of SMP protocols with and without public randomness,
respectively. The maxmin fluctuation results in a bottleneck for
private-coin protocols and the minmax for public-coin protocols.        
To obtain our lower bounds, we  generalize these and define the notion
of \emph{semimaxmin} decoupled chi-square fluctuation, which
interpolates between the maxmin and minmax fluctuations and captures
the setting of \emph{limited} public randomness.
\smallskip

In order to do so, we first recall the definition of
perturbations around a fixed distribution $\q\in\distribs{\ab}$. 
\begin{definition}[{\cite[Definition IV.4]{ACT:18:IT1}}]
Consider $0<\dst<1$, a family of distributions $\cP=\{\p_z, z\in \cZ\}$, and a distribution $\zeta$ on
$\cZ$. The pair $\cP_\zeta=(\cP, \zeta)$ is an \emph{almost $\dst$-perturbation
(around $\q$)} if
\[
\probaOf{\totalvardist{\p_Z}{\q}\geq \dst}\geq \alpha,
\]
for some $\alpha\geq 1/10$. We denote the set of all almost
$\dst$-perturbations by $\Upsilon_\dst$. Moreover, for $\alpha=1$ we
refer to $\cP$ as a \emph{perturbed family}. 
\end{definition}
For a channel $W\colon [\ab]\to\cY$, $z\in\cZ$, and a symbol
$y\in\cY$, we denote by $\q^W$ the distribution on $\cY$ induced
by $\q$ and $W$ (so that $\q^W(y) = \sum_{x\in[\ab]} W(y\mid
x)\q(x)$), and let $\delta_z^W(y) \eqdef
(\p_z^W(y)-\q^W(y))/\q^W(y)$. Also, for a family of channels $\cW$, denote by $\overline{\cW}$ its convex hull.
We now recall the definition of
decoupled chi-square fluctuation, and provide an operational meaning
for it.  
\begin{definition}[{\cite[Definition IV.3]{ACT:18:IT1}}]
  \label{def:induced:fluctuation} Consider a perturbed family
$\cP=\{\p_z: z\in \cZ\}$ and a family of channels
$\cW$. The 
$\ns$-fold \emph{induced decoupled chi-square fluctuation} of $\cP$
for $W^\ns \in \cW^\ns$ is given by
\[
\chisquaredec\left(W^\ns \mid \cP\right) \eqdef \ln \bE{ZZ^\prime}{\exp\left(\sum_{i=1}^\ns \dotprod{\delta_Z^{W_i}}{\delta_{Z'}^{W_i}}\right)}\,,
\]
where $\dotprod{\delta_z^W}{\delta_{z'}^W} = \bE{Y\sim \p^W}{\delta_z^W(Y)\delta_{z'}^W(Y)}$. 
\end{definition}
It was shown in previous work that
$\chisquaredec\left(W^\ns \mid \cP\right)$ is an upper bound on the chi-square
distance over the $\ns$ channel output distributions induced by the
almost $\dst$-perturbation, and $\q$; in particular, for any
testing protocol to be successful, this quantity must be bounded away
from zero.   
After these definitions, we are now in position to introduce the main
tool underlying our randomness tradeoff lower bound, the new notion
of \emph{semimaxmin} fluctuation: 
\begin{definition}[Semimaxmin Chi-square Fluctuation]
    \label{def:minmax:maxmin:induced:fluctuation} For a family of
channels $\cW$ and $\pbits \in \N$, the \emph{$(\ns, \dst,\pbits)$-semimaxmin decoupled chi-square
fluctuation for $\cW$} is given by
\[
\ouchisquaredec(\cW^\ns, \dst,\pbits) \eqdef \sup_{\substack{\cW_\pbits \subseteq \bar{\cW}^\ns \\ \abs{\cW_\pbits} \leq 2^\pbits}}\inf_{\cP_\zeta\in \Upsilon_\dst}\,\bEE{ \chisquaredec\left(W^\ns \mid \cP_\zeta\right)\land 1}\,,
\]
where the supremum is over all multisets $\cW_\pbits$ of $\overline{\cW}^\ns$ of size at most $2^\pbits$, the infimum is over all almost $\dst$-perturbations $\cP_\zeta$, and the expectation over the uniform choice of $W^\ns$ from $\cW_\pbits$. 
\end{definition}
One may observe that when $\pbits = 0$ and $\pbits = \infty$, respectively, replacing the expectation by a supremum yields the maxmin and minmax formulations from previous work. Here, we consider instead an inner expectation, as it makes it easier to bound the resulting quantity in practice~--~while making the proof of~\cref{l:ic_testing_fluctuation_bound_semiprivate} only slightly more technical. Note that in the definition we take a supremum over $2^\pbits$ choices of $W^\ns$ to capture the fact that there are $\pbits$ public bits which determine the distribution over the channels. If only $\pbits$ bits of public randomness are available, we will show that any test using channels from $\cW$ will err with large constant probability if the above quantity 
$\ouchisquaredec(\cW^\ns, \dst,\pbits)$ 
is upper bounded by a sufficiently small constant.
\begin{lemma}[Semimaxmin decoupled chi-square fluctuation bound for testing]
  \label{l:ic_testing_fluctuation_bound_semiprivate}
  For $0<\dst <1$, $\pbits\in\N$, and a $\ab$-ary reference distribution $\p$, the sample complexity $\ns=\ns(\ab,\dst,\pbits)$ of
$(\ab, \dst)$-identity testing with $\pbits$ bits of public randomness using $\cW$ 
must satisfy
\begin{equation}
\ouchisquaredec(\cW^\ns, \dst,\pbits)\geq c\,,
\end{equation}
for some constant $c>0$ depending only on the probability of error.
\end{lemma}
\begin{proof}
The proof uses Le Cam's two-point method. Consider an almost $\dst$-perturbation $\cP_\zeta$: we note first that, due the use of private coins, the effective channel used by each user is a convex
combination of channels from $\cW$, namely it is a channel from
$\overline{\cW}$. Thus, when $X^\ns$ has distribution either $\p^\ns$
and $\p_z^\ns$, respectively, $Y^\ns$ has distribution $\p^{W^\ns}$
and $\p_z^{W^\ns}$ with $W^\ns\in \overline{\cW}^\ns$. The public randomness then allow the users to jointly sample from any distribution on $\overline{\cW}^\ns$ which can be sampled by $\pbits$ independent unbiased bits, that is from any uniform distribution on a multiset $\cW_\pbits \subseteq \overline{\cW}^\ns$ of size (at most) $2^\pbits$.

Now, for every choice of channels $W^\ns = (W_1, \dots, W_\ns)\in\overline{\cW}^\ns$,
by Pinsker's inequality and the concavity of logarithm,
\[
\totalvardist{\expect{\p_Z^{W^\ns}}}{\p^{W^\ns} }^2 \leq \frac{1}{2} \kldiv{\expect{\p_Z^{W^\ns}}}{\p^{W^\ns}}
\leq \frac{1}{2} \ln\Paren{ 1+\chisquare{\expect{\p_Z^{W^\ns}}}{\p^{W^\ns}} }\,.
\] 
Also, we have the trivial bound $\totalvardist{\expect{\p_Z^{W^\ns}}}{\p^{W^\ns} }^2\leq 1$.
Fix any multiset $\cW_\pbits\subseteq \overline{\cW}^\ns$. Over the uniformly random choice of $W_U^\ns \in\cW_\pbits$ (using the public randomness $U$), we then have using the concavity of square roots,
\begin{align*}
\bE{U}{ \totalvardist{\expect{\p_Z^{W_U^\ns}}}{\p^{W_U^\ns}}}^2 
&\leq \bE{U}{ 1\land \sqrt{\frac{1}{2}  \ln\Paren{ 1+\chisquare{\expect{\p_Z^{W_U^\ns}}}{\p^{W_U^\ns}} } } }^2 \\
&\leq \bE{U}{ 1\land \frac{1}{2}\ln\Paren{ 1+\chisquare{\expect{\p_Z^{W_U^\ns}}}{\p^{W_U^\ns}} } } \,.
\end{align*}
We then bound the right-side further using~\cite[Lemma III.V]{ACT:18:IT1}
with $\theta$ replaced by $z$, $Q_\vartheta^\ns = \p_z^{W_U^\ns}$ and
$P_i =\p^{W_U^\ns}$ to get
\begin{align*}
\bE{U}{ 1\land \ln\Paren{ 1+\chisquare{\expect{\p_Z^{W_U^\ns}}}{\p^{W_U^\ns}} } }
&\leq \bE{U}{1\land \ln \bE{ZZ^\prime}{\prod_{i=1}^\ns (1+H^U_i(Z,Z^\prime))} }
\\
&\leq \bE{U}{1\land \ln  \bE{ZZ^\prime}{e^{\sum_{i=1}^\ns H^U_i(Z,Z^\prime)}} } 
\\
&= \bE{U}{1\land \chisquaredec\left(W_U^\ns \mid \cP_\zeta\right) },
\end{align*}
since $H^U_i(Z,Z^\prime)=\dotprod{\delta_Z^{W_{U,i}}}{\delta_{Z'}^{W_{U,i}}}$. That is, we have\footnote{Dropping the constant $1/2$ for simplicity of the resulting bound.}
\begin{align}
\bE{U}{ \totalvardist{\expect{\p_Z^{W_U^\ns}}}{\p^{W_U^\ns}} }^2 \leq \bE{U}{ 1\land \chisquaredec\left(W_U^\ns \mid \cP_\zeta\right) }.
\label{e:chisquare_ttlvar}
\end{align}
Consider an $(\ns, \dst)$-test $\Tester$ using a public-coin protocol. Denote by
$U$ the public randomness and by $Y_1, \dots, Y_\ns$ the messages from
each user and by $\cZ_0$ the set of $z$ such that
$\totalvardist{\p_z}{\p}\geq \dst$. Since $\cP_\zeta$ is an almost $\dst$-perturbation,
$\probaOf{Z\in \cZ_0}\geq \alpha\geq1/10$. Also, for the test $\Tester$ we have
$\probaDistrOf{X^\ns\sim \p^\ns}{\Tester(U, Y^\ns)=1}\geq 11/12$ and
$\probaDistrOf{X^\ns\sim \p_z^\ns}{\Tester(U, Y^\ns)= 0}\geq 11/12$ for every $z\in \cZ_0$.
Thus, we obtain
\[
\frac 12 \probaDistrOf{X^\ns\sim \p^\ns}{\Tester(U, Y^\ns)= 1} +
\frac 12 \probaDistrOf{X^\ns\sim \expect{\p_Z^\ns}}{\Tester(U, Y^\ns)= 0} \geq \frac{11(1+\alpha)}{24}\geq \frac {121}{240},
\]
where the last inequality relies on the fact that $\alpha\geq 1/10$. Equivalently, 
\begin{align}
\frac 12 \probaDistrOf{X^\ns\sim \p^\ns}{\Tester(U, Y^\ns)\neq 1} +
\frac 12 \probaDistrOf{X^\ns\sim \expect{\p_Z^\ns}}{\Tester(U, Y^\ns)\neq
0} \leq \frac {119}{240}\,.
\label{e:bayes_error_bound2}
\end{align}
An important remark here is that the distribution of $W_U^\ns$ (that is, the
choice of $\cW_\pbits \subseteq \overline{\cW}^\ns$) does not depend on
$\cP_\zeta$.
The left-hand-side of~\cref{e:bayes_error_bound2} above coincides with the Bayes error for test $\Tester$ for the simple binary hypothesis testing problem of $\expect{\p_Z^{W_U^\ns}}$ versus $\bE{U}{\p^{W_U^\ns}}$, which must be at least
\[
\frac 12 \left( 1 - \bE{U}{ \totalvardist{\expect{\p_Z^{W_U^\ns}}}{\p^{W_U^\ns}} }\right).
\]
 Thus, we can find 
$\cW_\pbits$ 
such that for $W_U^\ns$ distributed uniformly on $\cW_\pbits$ and any almost $\dst$-perturbations $\cP_{\zeta}$
\[
\bE{U}{ \totalvardist{\expect{\p_Z^{W_U^\ns}}}{\p^{W_U^\ns}} }\geq\frac 1 {120}\,,
\]
which along with~\cref{e:chisquare_ttlvar}
yields
\begin{align}
\bE{U}{ 1\land \chisquaredec\left(W_U^\ns \mid \cP_\zeta\right) }\geq c,
\label{e:semimaxmin_chisquare_bound}
\end{align}
where $c= 1/14400$. The result follows upon taking minimum over all almost $\dst$-perturbations $\cP_{\zeta}$ and the maximum over all multisets $\cW_\pbits\in \overline{\cW}^\ns$ of size at most $2^\pbits$.
\end{proof}

In view of~\cref{l:ic_testing_fluctuation_bound_semiprivate}, it then suffices to come up with a particular reference distribution $\q$ of our choosing, and, for any type of constraint $\cW$, to upper bound $\ouchisquaredec(\cW^\ns, \dst,\pbits)$ as a function of $\ab,\dst,\pbits$ and (some quantity of) $\cW$. To do so, recalling the definition of semimaxmin decoupled chi-square fluctuation (\cref{def:minmax:maxmin:induced:fluctuation}), it suffices to do the following: for each fixed $\cW_\pbits \subseteq \overline{\cW}^\ns$ of size at most $2^\pbits$, construct an almost $\dst$-perturbation $\cP_\zeta=(\cP, \zeta)$ around our $\q$ such that $\bEE{ \chisquaredec\left(W^\ns \mid \cP_\zeta\right)}$ is small enough.
As previously mentioned, we will choose our reference distribution 
$\q$ to be the uniform distribution $\uniformOn{\ab}$. Our almost perturbations will consist of ``small local perturbations'' around uniform, and be of the form
\begin{equation}\label{eq:paninski:pertubation}
    \p_Z = \frac{1}{\ab}( {1+Z_1\dst} , {1-Z_1\dst}, \dots, {1+Z_{\ab/2}\dst} , {1-Z_{\ab/2}\dst}  )\,,
\end{equation}
where $Z$ is drawn for a suitably chosen distribution $\zeta$ on $\R^{\ab/2}$. Note that taking $\zeta$ to be uniform on $\{-1,1\}^{\ab/2}$, we retrieve the ``Paninski construction''~\cite{Paninski:08}, widely used to prove lower bounds in the centralized, unconstrained setting.
Unfolding the definition of decoupled
chi-square perturbation, the form chosen in~\eqref{eq:paninski:pertubation} for our perturbation then naturally leads to  the following channel-dependent matrix $H(W)$, which will guide the choice of the ``worst possible mixture $\zeta$
over $\cZ$'' for a given family of channels. For each channel $W\in\cW$, let the $(\ab/2)$-by$(\ab/2)$ positive semidefinite matrix $H(W)$ be defined as
\begin{equation}\label{eq:def:barH} 
H(W)_{i_1, i_2} \eqdef
\sum_{y\in \cY}\frac{(W( y\mid 2i_1-1) -  W( y\mid 2i_1))(W( y\mid 2i_2-1)
  -W( y\mid 2i_2))}{\sum_{x\in [\ab]} W( y\mid x)}, \;\; i_1,
   i_2\in[\ab/2]\,.
\end{equation}
This matrix will, loosely speaking, capture the ability of channel $W$ to discriminate between even and odd
inputs, and thus to distinguish the reference uniform distribution from such a mixture of perturbed distributions. Our bounds will rely 
on the nuclear norm $\norm{H(W)}_\ast$ of the matrix $H(W)$. In
effect, our results characterize the \emph{informativeness} of a
channel $W$ for testing in terms of the nuclear norm of
$H(W)$. Channels with larger nuclear norms provide more information,
and the channel constraints impose a bound on the nuclear norms, which
leads to our result: 
\begin{theorem}
  \label{theo:semimaxmin}
Given $\ns\in \N$, $\dst\in (0,1)$, $\pbits\in\N$, for a channel family $\cW$ the
$(\ns,\dst,\pbits)$-semimaxmin chi-square fluctuation is bounded as
\begin{equation*}
\ouchisquaredec(\cW^\ns, \dst, \pbits) 
=
O\left( \frac{\ns^2\dst^4}{\ab^3}\cdot 2^{\pbits} \cdot \max_{W\in \cW}\, \norm{H(W)}_\ast^2\right)\,,
\end{equation*}
whenever
\begin{equation}
\ns \leq \frac{\ab^{3/2}}{C\dst^2 2^{\pbits/2} \max_{W\in\cW}\, \norm{H(W)}_\ast}\,,
\end{equation}
where $C>0$ is a constant.
\end{theorem} 
\noindent The proof of this theorem is quite technical, and is provided in~\cref{app:semimaxmin:lb}. We here give an outline of the argument.
\begin{proofof}{\cref{theo:semimaxmin} (Sketch)}
In view of the discussion above, we would like, given any multiset $\cW_\pbits$ of $2^\pbits$ $\ns$-fold channels $W^\ns$, to design a suitable distribution for our perturbation $Z$ which ``fools'' all (or most) of the $2^\pbits$ channels. Loosely speaking, we would like to construct a distribution for which (informally) most of variance  falls in subspaces corresponding to small eigenvectors for a large fraction of the matrices $H(W_i)$. To do so, we proceed along the same lines as the proof of~\cite[Theorem IV.18]{ACT:18:IT1} (hereafter denoted $(\star)$), reducing the problem to finding a distribution of the perturbation vector $Z$ such that, for any fixed (multi)set $\cW_\pbits \subseteq \cW^\ns$ of size at most $2^\pbits$, the expectation
\[
\bE{W^\ns}{ \ln \bE{ZZ'}{ e^{ \frac{\beta^2\ns^2\dst^2}{\ab} Z^\transp \bar{H}(W^\ns) Z' } } }
\]
(where $\beta>0$ is a constant, and $\bar{H}(W^\ns) \eqdef \frac{1}{\ns}\sum_{i=1}^\ns H(W_i)$), is small. Using a similar argument, it suffices to find a matrix $V$ such that (i)~$\norm{V}_F^2\gtrsim \ab$, (ii)~each row of $V$ has 2-norm at most 1, and (iii)~the average (over $W^\ns\in\cW_{\pbits}$) Frobenius norm $\bE{W^\ns}{\norm{V^\transp \overline{H}(W^\ns) V}_F}$ is small. 

Since all the matrices $\bar{H}(W^\ns)$ (and therefore all $V^\transp \bar{H}(W^\ns) V$'s) are symmetric positive semi-definite matrices, one can then show that
\begin{equation}\label{ineq:frobenius:psd}
    \frac{1}{2^\pbits}\norm{V^\transp \Paren{\sum_{W^\ns\in\cW_{\pbits}} \bar{H}(W^\ns)} V }_F^2 \geq \bE{W^\ns}{\norm{V^\transp \bar{H}(W^\ns) V}_F}^2\,.
\end{equation}
Using a construction from $(\star)$ applied to $\tilde{H}(\cW_{\pbits}) \eqdef \sum_{W^\ns\in\cW_{\pbits}} \bar{H}(W^\ns)$, we obtain a matrix $V$ satisfying the above conditions~(i) and~(ii), and such that we have the following analogue of~(iii): 
\begin{equation}
      \norm{V^\transp\tilde{H}(\cW_{\pbits}) V}_F^2 \lesssim \frac{1}{\ab}\norm{\tilde{H}(\cW_{\pbits})}_\ast^2\,.
\end{equation}
Combining this inequality with~\eqref{ineq:frobenius:psd} and the triangle inequality, this leads to
\begin{equation}\label{eq:semimaxmin:bound:tildeH}
    \bE{W^\ns}{\norm{V^\transp \bar{H}(W^\ns) V}_F}^2 \lesssim \frac{2^\pbits}{\ab}\max_{W\in \cW} \norm{H(W)}_\ast^2\,.
\end{equation}
From~\eqref{eq:semimaxmin:bound:tildeH}, we can finally derive the desired bound in a manner analogous to the end of $(\star)$. This is however not entirely immediate, as (by our very construction), we can only guarantee small Frobenius norms and spectral radius \emph{on average} for the $V^\transp \bar{H}(W^\ns) V$'s. The original argument of~$(\star)$, however, crucially requires during its last step a pointwise guarantee; to conclude, we thus must resort to a careful averaging argument over these spectral radii to ensure \emph{most} of them are under control, and handle the small remaining ``bad'' fraction separately. More specifically, this last part hinges on the inner $\min$ in the definition of semimaxmin fluctuation: when bounding the quantity $\shortexpect_{W^\ns}[ \chisquaredec\left(W^\ns \mid \cP_\zeta\right)\land 1]$ in the end, we control the pointwise contribution of the ``good'' $W^\ns$'s via the term $\chisquaredec\left(W^\ns \mid \cP_\zeta\right)$ (which we show is then $\ll 1$), and the contribution of the ``bad'' $W^\ns$'s via the term $1$ (which, while large, is weighted by the fraction of ``bad'' channels, which is itself small enough).
\end{proofof}

\subsection{Communication-Constrained and LDP Testing} 
We now instantiate the general lower bound result established in the previous section to the two specific settings we consider, communication and local privacy constraints. For communication-constrained and LDP channels the nuclear norms of the $H$ matrices can be uniformly bounded as follows.
\begin{lemma}[{\cite[Lemmas V.1 and V.5]{ACT:18:IT1}}]
\label{l:norm:bounds}
For $\numbits \geq 1$, and $\priv \in(0,1]$,
$\max_{W\in\cW_\numbits}\, \norm{H(W)}_\ast\le 2^\numbits$ and
$\max_{W\in\cW_\priv}\, \norm{H(W)}_\ast = O(\priv^2)$.
\end{lemma}
Using these bounds, we readily obtain our sample complexity results for both communication-constrained and LDP channels. 
\begin{theorem}
\label{c:testing:semi:public:private:lbits:communication}
For $0<\dst <1$ and $\numbits,\pbits \in\N$, the sample complexity of $(\ab, \dst)$-uniformity
testing  with $\pbits$ bits of public randomness using $\cW_\numbits$ is at least
\[
 \bigOmega{\frac{\sqrt{\ab}}{\dst^2}\sqrt{\frac{\ab}{2^{\numbits}}\lor 1}\sqrt{\frac{\ab}{2^{\pbits+\numbits}}\lor 1} }\,.
\]
\end{theorem}

\begin{theorem}
\label{c:testing:semi:public:private:lbits:ldp}
For $0<\priv <1$, and $\pbits\in\N$ the sample complexity of $(\ab, \dst)$-uniformity
testing  with $\pbits$ bits of public randomness using $\cW_\priv$ is at least
\[
 \bigOmega{ \frac{\sqrt{\ab}}{\dst^2} \frac{\sqrt{\ab}}{\priv^2}  \sqrt{\frac{\ab}{2^{\pbits}}\lor 1}}\,.
\]
\end{theorem}
Indeed, from~\cref{l:ic_testing_fluctuation_bound_semiprivate}, we get that $\ouchisquaredec(\cW^\ns, \dst, \pbits) $ must be lower bounded by a constant for $\ns$ samples to be sufficient for testing. Plugging in the bounds from~\cref{l:norm:bounds} in~\cref{theo:semimaxmin} yields the two above results.

\clearpage
\bibliographystyle{alpha}
\bibliography{randomnesstradeoffs}

\appendix
\section{Spectrum of outer products result}\label{app:codebook-construction}
In this appendix we prove~\cref{t:super_sets} and~\cref{l:additivity_tails}. \cref{t:super_sets} is restated below:
\begin{theorem}[Spectrum of outer products]\label{t:super_sets:restated}
For $n\in \N$, there exist constants $c_0\in \N$, $c_1,c_2\in
(0,1)$ and vectors $u_1, \dots, u_{m_0}\in \{0,1\}^n$ with $m_0 =2^{c_0}n$ such
that for every $\cJ \subseteq [m_0]$ with $|\cJ|\geq (1-c_1)m_0$ we must
have
\[
\lambda_{\min}\bigg(\frac 1 {|\cJ|}\sum_{j\in \cJ}u_ju_j^\transp\bigg)\geq c_2. 
\]
\end{theorem}

Consider random, independent binary vectors $V_1, \dots,V_{m_0}\in\{0,1\}^n$, with each $V_i$ drawn uniformly from the set of all
binary vectors of length $n$. We establish~\cref{t:super_sets:restated} using probablistic argument. It would be enough to show that:

\[
\probaOf{\exists\, \cJ \subseteq [m_0], |\cJ|\geq (1-\theta)m_0\text{ s.t. }
	\lambda_{\min}\left(\frac 1 {|\cJ|}\sum_{i\in\cJ}V_iV_i^\transp\right)< c_2} < 1.
\]

First, for any $\cJ$ with $|\cJ| = m \ge (1-\theta) m_0$, we will derive an exponential upper bounds for the probability, 
\[
	\probaOf{\lambda_{\min}\left(\frac 1 {|\cJ|}\sum_{i\in\cJ}V_iV_i^\transp\right) < t} . 
\]

Without loss of generality, we can assume $\cJ = [m]$. Since 
\[
	\lambda_{\min}\left(\frac 1 {|\cJ|}\sum_{i\in\cJ}V_iV_i^\transp\right) = \min_x\{ \frac{\frac1{m} \sum_{j=1}^m x^\transp V_jV_j^\transp x }{\norm{x}_2^2}\},
\]
we first establish an exponential upper bound for
\[
	\probaOf{\frac1{m} \sum_{j=1}^m x^\transp V_jV_j^\transp x < t \norm{x}_2^2}.
\]
We derive this bound using a general anti-concentration bound for
subgaussian random variables, which may be of independent interest.
\begin{theorem}[An anti-concentration bound]
  \label{t:anti-concentration}
Consider independent random variables  $Y_1, \dots, Y_m$ such that each
$Y_i$ is zero-mean and subgaussian with variance
parameter $\sigma^2$, \ie{} $\expect{e^{\lambda Y_i}}\leq
e^{\lambda^2\sigma^2/2}$ for all $\lambda\in \R$. Suppose further that, for all $i$,
$\expect{Y_i^2} \geq \eta \sigma^2$ for some $\eta\in(0,1)$. 
Then, there exist positive constants $c_1$ and $c_2$ such that 
for every $\mu \in \R$, 
\[
\probaOf{\frac 1 m\sum_{i=1}^m(Y_i+\mu)^2\geq
c_1\eta^2\Big( \min_{1\leq i\leq m}\expect{Y_i^2}+\mu^2\Big)}\geq 1- \exp(-c_2 m\eta^4).
\]
\end{theorem}
To prove this result, we take recourse to the following ``clipped-tail'' version
of Hoeffding bound, which allows us to obtain exponential 
anti-concentration bounds using anti-concentration bounds. 
\begin{lemma}[Clipped-tail Hoeffding bound]\label{l:CTH}
For $t>0$, let $X_1, \dots, X_m$ be nonnegative, independent random variables satisfying
\[
\probaOf{X_i\geq t}\geq \alpha, \qquad 1\leq i\leq m,
\]
Then, 
\[
\probaOf{\frac 1 m\sum_{i=1}^m X_i \geq \frac{t\alpha}{2}}\geq 1
-\exp\left(-m\frac{\alpha^2}2\right). 
\]
\end{lemma}
\begin{proof}
Since $X_i$s are nonnegative, $\sum_{i=1}^m X_i\geq
t\sum_{i=1}^m \indic{X_i>t}$. It follows that
\[
\probaOf{\frac 1 m\sum_{i=1}^m X_i \leq \frac{t\alpha}{2}}=
\probaOf{\frac 1 m\sum_{i=1}^m \indic{X_i>t} \leq \frac{\alpha}{2}},
\]
where the right-side is bounded above further by $\exp(-m\alpha^2/2)$
using Hoeffding's inequality and the assumption of the lemma.
\end{proof}
We use this bound to now complete the proof
of~\cref{t:anti-concentration}. 

\begin{proofof}{\cref{t:anti-concentration}} Let $Y$ be zero-mean and subgaussian with
variance parameter $\sigma^2$. Then, for $X=Y+\mu$, 
we get $\expect{X^4} \leq 8\expect{Y^4} + 8\mu^4$. Also, since $Y$ is
subgaussian with variance parameter $\sigma^2$, it is easy to show
that $\expect{Y^4}\leq 8\sigma^2$, whereby we get
$\expect{X^4}\leq 64\sigma^4 + 8\mu^4$.
Since by our assumption
$\expect{X^2}=\expect{Y^2}+\mu^2\geq \eta \sigma^2+\mu^2$, it
follows that $\expect{X^2}^2\geq \eta^2\sigma^4+\mu^4$.
Upon combining this with the previous bound, we obtain
$\expect{X^4}\leq \frac{64}{\eta^2}\expect{X^2}^2$. We now invoke the Paley--Zygmund inequality to get
\[
\probaOf{X^2\geq \frac 1 2\,(\expect{Y^2}+\mu^2)}\geq \frac {\eta^2}{256}.
\]
Finally, an application of~\cref{l:CTH} yields
\[
\probaOf{\frac 1m\sum_{i=1}^m (Y_i+\mu)^2\geq \frac{\eta^2}{1024}\,
\Big(\min_{1\leq i\leq m}\expect{Y_i^2}+\mu^2\Big)}
\geq 1-\exp\left(-\frac{m\eta^4}{256^2}\right),
\]
which completes the proof.
\end{proofof}

\begin{proofof}{\cref{t:super_sets:restated}}
Let $\one$ be the all one vector in $R^n$. We apply~\cref{t:anti-concentration} to $Y_i=x^\transp V_i
-(\one^\transp x)/2$, $1\leq i\leq m$, with $\mu=(\one^\transp x)/2$. Note
that the $Y_i$'s are zero-mean, and by Hoeffding's lemma, they are subgaussian with
variance parameter $\normtwo{x}^2/4$. Furthermore, it is easy to verify that
$\expect{Y_i^2}=\normtwo{x}^2/4$. Thus, the condition
of~\cref{t:anti-concentration} holds with $\eta=1$, which gives
\begin{align}
\probaOf{\frac 1m\sum_{i=1}^m (x^\transp V_i)^2
\geq c_1\,\frac{\normtwo{x}^2+(\one^\transp x)^2}4
}\geq 1-\exp\left(-c_2m\right).
\label{e:single-vector}
\end{align}
Denote by $A_m$ the random matrix $\frac{1}{m}\sum_{i=1}^mV_iV_i^\transp.$
Our goal is to bound $\lambda_{\min}(A_m)$. 
It will be convenient to introduce a new norm $\norm{\cdot}_{\star}$
on $\R^n$: for $x\in \R^n$,
\[
\norm{x}_{\star} \eqdef \sqrt{\norm{x}_2^2 + (\mathbf{1}^\transp
x)^2}.
\]
Clearly, $\norm{\cdot}_{\star}$ is a norm, as
$\norm{x}_{\star} = \normtwo{L(x)}$ where $L(x) \eqdef
(x_1,\dots,x_n, \sum_{i=1}^n x_i)\in\R^{n+1}$ is linear. 

Now, if we can find an $x$ such that $x^\transp A_m
x< \lambda \normtwo{x}^2$, then $y=x/\norm{x}_\star$ has
$\norm{y}_\star=1$ and satisfies $y^\transp A_m
y< \lambda$. Therefore, 
\begin{align}
\probaOf{\min_{x:
\normtwo{x}=1}x^\transp A_m x <\lambda}
\leq 
\probaOf{\min_{y: \norm{y}_\star =1}y^\transp A_m y<\lambda}
\label{e:norm-change}
\end{align}
We use~\cref{e:single-vector} to obtain this bound, together with an appropriate netting argument. Specifically, let
$\mathcal{N}$ be a $\delta$-net of the sphere
$\setOfSuchThat{y\in \R^n }{ \norm{y}_\star =1 }$ in the norm
$\norm{\cdot}_{\star}$. We can find such a net with $|\mathcal{N}|\le
(1+\frac{2}{\delta})^n \leq e^{2n/\delta}$ (see, \eg~\cite[Lemma~4.16]{Pisier:99}), which is the net we use. By a union bound applied to~\cref{e:single-vector},
we get
\begin{align}
\probaOf{\min_{x\in \cN}x^\transp A_m x
< c_1\,\frac{\normtwo{x}^2+(\one^\transp x)^2}4
}<\exp\left(\frac{2n}{\delta}-c_2m\right).
\label{e:net}
\end{align}
We bound $y^\transp A_m y$ for
a $y$ with $\norm{y}_\star=1$ by relating it to $x^\transp A_m x$ for
a vector $x\in \cN$ such that $\norm{x-y}_\star$. While this is he
standard netting argument, there is added complication since we need
to work with the norm $\norm{\cdot}_\star$.

In particular, for a
$y$ such that $\norm{y}_\star$ consider an $x\in \cN$ satisfying $\norm{x-y}_{\star}\le \delta$. 
Denoting $z\eqdef y-x$, we decompose $z=z_{\parallel} + z_{\perp}$, where
$z_{\parallel}\in \text{span}_{\R}(\one)$, and
$z_{\perp}^\transp \one=0$. By definition,
$z_{\parallel}=\frac{(z^\transp \one)}n \one$
and $z_\perp=z-z_\perp$. 
Using the inequality $(a+b)^2 \geq a^2/2-b^2$, for every $i\in [m]$ we have
\begin{align*}
(V_i^\transp y)^2 = (V_i^\transp x + V_i^\transp
z)^2 \geq \frac{1}{2}(V_i^\transp x)^2 - (V_i^\transp
z)^2 \geq \frac{1}{2}(V_i^\transp x)^2 - 2(V_i^\transp z_\parallel)^2
- 2(V_i^\transp z_\perp)^2.  
\end{align*}
Summing over $i$ and using the expression for $z_\parallel$, we get
\begin{align}
y^\transp A_m y\geq \frac 12 \cdot x^\transp A_m x- \frac{2(z^\transp \one)^2}{n^2}\cdot (\one^\transp
A_m \one) - 2 (z_\perp^\transp A_m z_\perp). 
\label{e:intermediate1}
\end{align}
To proceed further, we derive bounds for random variables 
$(\one^\transp A_m \one)$ and $(z_\perp^\transp A_m z_\perp)$. 
For the first term, we can show 
\begin{align}
\probaOf{(\one^\transp A_m \one)>5n^2}\leq 2\exp(-m/2).
\label{e:intermediate2}
\end{align}
We provide a proof at the end. For the second term, we observe that
\[
z_\perp^\transp A_m z_\perp=
\frac 1m \sum_{i=1}^m(z_\perp^\transp V_i)^2
= \frac 1m \sum_{i=1}^m\left(z_\perp^\transp \left(V_i-\frac 12 \cdot \one\right)\right)^2. 
\]
Denote by $\overline{V}_i$ a random variable which takes values $1/2$
and $-1/2$ with equal probabilities, and by $\overline{A}_m$ the
random matrix
$(1/m)\sum_{i=1}^m \overline{V}_i\overline{V}_i^{\,\transp}$, we get
\[
z_\perp^\transp A_m z_\perp\leq \lambda_{\max}(\overline{A}_m)\normtwo{z_\perp}^2.
\]
The next result, whose proof is standard and will be given later, 
provides a bound for $\lambda_{\max}(\overline{A}_m)$. 
\begin{lemma}\label{l:max_eigenvalue}
There exist constants $c_2, c_3$ such that
\[
\probaOf{\lambda_{\max}(\overline{A}_m)> c_2}\leq \exp\left(c_3n-\frac m2\right).
\]
\end{lemma}
\noindent This result, together with~\cref{e:intermediate1} and~\cref{e:intermediate2}, yields
\begin{align*}
&\lefteqn{\probaOf{\min_{y: \norm{y}_\star=1}y^\transp A_m y\geq t}}
\\
&\geq \probaOf{
\min_{x\in \cN} x^\transp A_m x\geq 2t+ 
20(z^\transp \one)^2+ 4c_2\normtwo{z_\perp}^2
}
-2\exp\left(- \frac m2\right)
- \exp\left(c_3n-\frac m2\right)
\\
&\geq 1- \probaOf{
\min_{x\in \cN} x^\transp A_m x\geq 2t+ 
c_4\delta^2
}
-2\exp\left(- \frac m2\right)
- \exp\left(c_3n-\frac m2\right)
\end{align*}
where we used
$\norm{z}_\star^2\geq \normtwo{z_\perp}^2+(z^\transp \one)^2$. 
We set $t=\delta^2$ and note that for any $x\in \cN$ we must
have $\norm{x}_\star\leq 1+\delta$. Therefore, 
\[
\probaOf{
\min_{x\in \cN} x^\transp A_m x< 2t+ 
c_4\delta^2
}
= \probaOf{\exists\, x\in \cN\text{ s.t. } 
 x^\transp A_m x< c_5\frac{\delta^2}{(1+\delta)^2} \norm{x}_\star^2}.
\]
Setting $\delta$ such that $c_5\delta^2(1+\delta)^2=c_2/4$, it follows
from~\cref{e:single-vector} that
\[
\probaOf{
\min_{x\in \cN} x^\transp A_m x< 2t+ 
c_4\delta^2
}\leq |\cN|\exp(-c_2m)\leq \exp\left(\frac{2n}{\delta} - c_2m\right).
\]
Upon combining the bounds above, we get
\[
\probaOf{\min_{y:\norm{y}_\star=1}y^\transp A_m y< \delta^2}\leq
 \exp\left(\frac{2n}{\delta} - c_2m\right)
+2\exp\left(- \frac m2\right)
+\exp\left(c_3n-\frac m2\right)
\]
where $\delta, c_2, c_3$ are constants. 
Recalling~\cref{e:norm-change}, we have obtained
\[
\probaOf{
\lambda_{\min}\left(\frac 1m\sum_{i=1}^mV_iV_i^\transp\right)< \delta^2}\leq
 \exp\left(c_6n - c_7m\right)
\]
Finally, by a union bound of all subsets of $[m_0]$ with size larger than $(1-\theta)m_0$, we get
\[
\probaOf{\exists\, \cJ \subseteq [m_0], |\cJ|\geq (1-\theta)m_0\text{ s.t. }
\lambda_{\min}\left(\frac 1 {|\cJ|}\sum_{i\in\cJ}V_iV_i^\transp\right)< \delta^2}\leq
 m_02^{m_0h(\theta)} \exp\left(c_6n - c_7m_0(1-\theta)\right),
\]
where $h(\cdot)$ denotes the binary entropy function, and we have used
the fact that the number of subsets of $[m_0]$ of cardinality greater
than $(1-\theta)m_0$, $\theta\in (0, 1/2)$,  is at most
$m_02^{m_0h(\theta)}$. The proof is completed by ensuring that the
exponent on right-side above is negative.
\end{proofof}

\noindent It only remains to prove~\cref{e:intermediate2,l:max_eigenvalue}, which we do next.
\begin{proofof}{\cref{e:intermediate2}}
Consider random variables $\xi_i\eqdef (V_i^\transp \one)$, $i\in [m]$. Note that $\expect{\xi_i}=n/2$ and each $\xi_i$ is subgaussian
with variance parameter $n/4$. Therefore,
$\probaOf{\frac 1m\sum_{i=1}^m\xi_i>n}\leq \exp(-mn/2)$. 
Furthermore, since $\expect{(\xi_i - n/2)^2}=n/4$, the random variable
$(\xi_i - n/2)^2-n/4$ is subexponential with parameter $4n$, which
gives
$\probaOf{
\frac 1m\sum_{i=1}^m(\xi_i-n/2)^2>17n/4}\leq \exp(-m/2)$. 
Thus, 
\begin{align*}
\probaOf{\frac 1 m\sum_{i=1}^m \xi_i^2> \frac 34\cdot n^2 + \frac
{17}4\cdot n}
&\leq
\probaOf{\frac 1m\sum_{i=1}^m\xi_i>n}+
\probaOf{\frac 1m\sum_{i=1}^m(\xi_i-n/2)^2>5n/4}
\\
&\leq 2\exp\left(-\frac m2\right),
\end{align*}
which leads to the claimed bound. 
\end{proofof}

\begin{proofof}{\cref{l:max_eigenvalue}}
For a fixed $x\in \R^n$, consider random variables
$\zeta_i \eqdef (\overline{V}^\transp x)$, $i\in[m]$. They are all zero-mean
and are subgaussian with variance parameter
$\normtwo{x}^2/4$. Furthermore, their second moment
$\expect{\zeta_i^2}$ equals $\normtwo{x}^2/4$. Therefore, the random
variable $\zeta_i^2-\normtwo{x}^2/4$ is subexponential with parameter
$4\normtwo{x}^2$, and we have
$\probaOf{\frac 1m \sum_{i=1}^m \zeta_i^2>\frac{17}4 \cdot \normtwo{x}^2}
\leq \exp\left(-\frac m2\right)$.

Next, consider a $\delta$-net $\cN_2$ of the unit ball under
$\normtwo{\cdot}$ of cardinality $|\cN_2|\leq e^{2n/\delta}$. For a
$y$ such that $\normtwo{y}=1$ and $y^\transp \overline{A}_m
y=\lambda_{\max}(\overline{A}_m)$, consider the $x\in \cN_2$ such that
$\normtwo{y-x}\leq\delta$.
Then, since $y^\transp \overline{A}_m y= x^\transp \overline{A}_m
x+2(y-x)^T\overline{A}_my$, we have
\[
\lambda_{\max}(\overline{A}_m)=y^\transp \overline{A}_m y\leq x^\transp \overline{A}_m
x+2\delta\lambda_{\max}(\overline{A}_m),
\]
which further gives
\[
(1-2\delta)\lambda_{\max}(\overline{A}_m)
\leq \max_{x\in \cN_2}x^\transp \overline{A}_m x.
\]
Also, every $x\in \cN_2$ satisfies $\normtwo{x}\leq 1+\delta$, a
and so, by the tail-probability bound for $\sum_{i=1}^m\zeta_i^2$ that
we saw above, we get $\probaOf{\frac
1m \sum_{i=1}^m \zeta_i^2>\frac{17(1+\delta)^2}4}  \exp\left(2n/\delta- m/2\right)$.
Therefore, we obtain
\[
\probaOf{\lambda_{\max}(\overline{A}_m)>\frac{17(1+\delta)^2}{4(1-2\delta)}}
\leq \exp\left(-\frac m2\right).
\]
In particular, we can set $\delta=1/4$ to get the claimed result with
$c_2=425/32$ and $c_3=8$.
\end{proofof}

We close with a proof of~\cref{l:additivity_tails}, which we recall below for easy reference.
\begin{lemma}[Additivity of tails~\cref{l:additivity_tails}, restated]
Let $a_1,\dots,a_m\geq 0$, and suppose $Y_1, \ldots, Y_m$ are
non-negative random variables with $\probaOf{Y_i \geq a_i} \geq c$ for
every $1\leq i\leq m$, for some $c\in(0,1)$. Then,
\[
\probaOf{Y_1+\ldots+Y_m \geq
c\cdot \frac{a_1+\ldots+a_m}{2}} \geq \frac{c}{2-c}\,.
\]	
\end{lemma}
		\begin{proof}
			Let $a_1,\dots,a_m\geq 0$ and $Y_1, \ldots, Y_m$ be as in the statement, and define 
			$Z_i \eqdef a_i \indic{Y_i \geq a_i}$ for $i\in[m]$. Then $Z_1, \ldots, Z_m$ satisfy the assumptions of the lemma as well, namely $\probaOf{Z_i\geq a_i}\geq c$.
Further,
			$\probaOf{Y_1+\ldots+Y_m \geq \alpha} \geq \probaOf{Z_1+\ldots+Z_m \geq \alpha}$ for all $\alpha$. Thus it suffices to prove the statement for the $Z_i$'s, which are supported on two points, which is what we do. 

			Let $Z \eqdef Z_1+\ldots+Z_m$. By the assumption, we have $\expect{Z} \geq c(a_1+\ldots+a_m)$, and 
			$0 \leq Z \leq a_1+\ldots+a_m$. 
			By Markov's inequality applied to $\sum_{i=1}^m a_i - Z \geq 0$, for $\gamma \in (0,1)$,
			\[
			\probaOf{ Z < \gamma c \sum_{i=1}^m a_i }  = \probaOf{ \sum_{i=1}^m a_i - Z > (1-\gamma c)\sum_{i=1}^m a_i }
			\leq \frac{ \sum_{i=1}^m a_i - \expect{Z}}{ (1-\gamma c)\sum_{i=1}^m a_i }
			\leq \frac{1-c}{1-\gamma c}
			= 1 - \frac{(1-\gamma)c}{1-\gamma c}\,.
			\]
			Taking $\gamma \eqdef 1/2$ yields the claim.
		\end{proof}

 \section{Miscellaneous: some useful lemmas}\label{app:useful:lemmas}
We provide in this appendix two simple results, mentioned in the preliminaries. We begin with a simple proposition, which allowed us throughout the paper on to assume that one can partition the domain $[\ab]$ into any number $L$ of equal-sized sets. Indeed, as shown below, when aiming to perform $(\ab,\dst)$-identity testing this can always be achieved at the cost of only a constant multiplicative factor in the distance parameter $\dst$ (and only requires private randomness, as well as knowledge of $\ab$ and $L$, from the $\ns$ users).
\begin{proposition}
    \label{prop:prelim:divisibility}
  Let $\ab, L\geq 1$ be two integers with $1\leq L\leq \ab$, and define $\ab' \eqdef L\clg{\ab/L}$. There exists an explicit mapping $\Phi_{\ab,L}\colon \distribs{\ab}\to\distribs{\ab'}$ such that
  (i)~the uniform distribution is mapped to the uniform distribution, \ie $\Phi_{\ab,L}(\uniform_\ab)=\uniform_{\ab'}$; and
  (ii)~distances are approximately preserved: for every $\p,\q\in\distribs{\ab}$,
  \[
    \totalvardist{\Phi_{\ab,L}(\p)}{\Phi_{\ab,L}(\q)} = \frac{\ab}{\ab'}\totalvardist{\p}{\q} \geq \frac{1}{2}\totalvardist{\p}{\q}\,.
  \]
  Further, there exists a randomized mapping $\Psi_{\ab,L}$ such that, for every $\p\in\distribs{\ab}$, $\Psi_{\ab,L}(X)\sim\Phi_{\ab,L}(\p)$ whenever $X \sim \p$.
\end{proposition}
\begin{proof}
    We define $\Phi_{\ab,L}$ as a mixture of the input and the uniform distribution on $[\ab]\setminus[\ab']$: for any $\p\in\distribs{\ab}$,
    $
        \Phi_{\ab,L}(\p) \eqdef \frac{\ab}{\ab'}\p + \frac{\ab'-\ab}{\ab'}\uniform_{[\ab]\setminus[\ab']}\,.
    $
    Recalling that $\ab \leq \ab' < \ab + L$, is immediate to verify that all the claimed properties hold.
\end{proof}

\noindent Applying the above with $L\eqdef 2^{\flr{\log\ab}}$, we in particular get the following:
\begin{corollary}
  Let $\ab \geq 1$ be any integer, and define $\ab' \eqdef 2^{\clg{\log\ab}} \in [\ab, 2\ab)$. There exists an explicit mapping $\Phi_{\ab}\colon \distribs{\ab}\to\distribs{\ab'}$ such that, for every $\p,\q\in\distribs{\ab}$,
  \[
    \frac{1}{2}\totalvardist{\p}{\q} \leq \totalvardist{\Phi_{\ab}(\p)}{\Phi_{\ab}(\q)} \leq \totalvardist{\p}{\q}\,.
  \]
  Further, there exists a randomized mapping $\Psi_{\ab}$ such that, for every $\p\in\distribs{\ab}$, $\Psi_{\ab}(X)\sim\Phi_{\ab}(\p)$ whenever $X \sim \p$.
\end{corollary}
\noindent In view of this corollary, we without loss of generality  can assume throughout that $\ab$ is a power of two.
 \section{Omitted proof: Deterministic Amplification}\label{app:deterministic:amplification}
In this appendix, we provide for completeness a proof of~\cref{theo:deterministic:expansion}, the ``deterministic error reduction'' lemma we used in the argument of~\cref{theo:upper:bounds:general}. The idea underlying this deterministic error reduction for $\textsf{RP}$ is well-known, and was introduced by Karp, Pippenger, and Sipser in 1985~\cite{KPS85}. The gist is to see the random string $\sigma$ as the index of a vertex is a $d$-regular graph on $2^\pbits$ vertices, and then run the algorithm on all $d$ neighbors of this random vertex $v_r$. If the graph is a good enough expander, doing so will ensure not all $d$ neighbors cause the algorithm to err. (For more on deterministic amplification for $\textsf{RP}$ (one-sided) and $\textsf{BPP}$ (two-sided) algorithms, as well as the related notion of exponential error amplification with \emph{few} extra random bits, see, \eg~\cite{CW89,CG89}, or~\cite[Sections 1.3.3 and 3.3]{HLW06}).

We begin by recalling some definitions and a useful lemma. Fix $n,d\in\N$ and $\lambda\geq 0$. We say that a $d$-regular graph $G=(V,E)$ on $n$ vertices with (normalized) adjacency matrix $A$ has \emph{spectral expansion} $\lambda$ if $\lambda(G)\leq \lambda$, where $\lambda(G) \eqdef \max(\dabs{\lambda_2},\dabs{\lambda_n})$ and $1\geq \lambda_1 \geq \lambda_2 \geq \dots \geq \lambda_n\geq -1$ are the eigenvalues of $A$.

\begin{theorem}[Expander Mixing Lemma]\label{theo:expander:mixing}
  Let $G=(V,E)$ be a $d$-regular graph on $n$ vertices with spectral expansion $\lambda$. Then, for every $S,T\subseteq V$,
  \[
      \dabs{ \frac{\dabs{e(S,T)}}{d n} - \frac{\dabs{S}}{n}\cdot \frac{\dabs{T}}{n} } \leq \lambda\sqrt{ \frac{\dabs{S}}{n}\cdot \frac{\dabs{T}}{n} }\,,
  \]
  where $e(S,T) = \setOfSuchThat{ (u,v)\in E }{ u\in S, v\in T }$.
\end{theorem}
\noindent We are now ready to prove~\cref{theo:deterministic:expansion}, restated below.
\begin{lemma}[Deterministic Amplification for One-Sided Error (\textsf{RP})] \label{theo:deterministic:expansion:restated}
For any $\pbits \in\N$ and $\eta,\gamma\in(0,1)$, there exist $d=d(\eta,\gamma)$ and (time-efficiently computable) functions $\pi_1,\dots,\pi_d\colon\{0,1\}^\pbits\to\{0,1\}^\pbits$ such that the following holds. 
Suppose $\cX_0\subseteq \cX$ and $A\colon \cX\times\{0,1\}^\pbits\to \Omega$ and $\cE\subseteq \Omega$ satisfy 
\begin{enumerate}[(i)]
  \item If $x\in\cX_0$, $\probaDistrOf{\sigma \sim\{0,1\}^\pbits}{A(x,\sigma)\in\cE} = 1$ \hfill(Perfect completeness)
  \item If $x\notin\cX_0$, $\probaDistrOf{\sigma \sim\{0,1\}^\pbits}{A(x,\sigma)\notin\cE} \geq 1-\eta$ \hfill(Low soundness)
\end{enumerate}
Then we have
\begin{enumerate}[(i)]
  \item If $x\in\cX_0$, $\probaDistrOf{\sigma \sim\{0,1\}^\pbits}{\forall i\in[d],\, A(x,\pi_i(\sigma))\in\cE} = 1$ \hfill(Perfect completeness)
  \item If $x\notin\cX_0$, $\probaDistrOf{\sigma \sim\{0,1\}^\pbits}{\exists i\in[d],\, A(x,\pi_i(\sigma))\notin\cE} \geq 1-\gamma$ \hfill(High soundness)
\end{enumerate}
Moreover, on can take $d=\tildeO{ \frac{\eta}{(1-\eta)^2\gamma} }$.
\end{lemma}
\begin{proof}
 Fix $A$ as in the statement, and let $G=(V,E)$ be a $d$-regular graph on $n\eqdef 2^\pbits$ vertices with spectral expansion $\lambda \leq (1-\eta)\sqrt{\gamma/\eta}$, for some $d$. We define $\pi_1,\dots,\pi_d$ as follows: fixing any labeling of the vertices of $G$, we see $r\in\{0,1\}^\pbits$ as a vertex $v_r\in V$ and let $\pi_1(r),\dots,\pi_d(r)$ be the labels of the $d$ neighbors of $v_r$ in $G$.
 
 To see why the claimed properties hold, first note that whenever $x\in\cX_0$, then as $A$ has one-sided error we have $A(x,\pi_i(\sigma))\in\cE$ for all $i$ with probability one. To establish the second item, fix $x\notin\cX_0$, and define $B_x\subseteq \{0,1\}^\pbits$ as the set of ``bad'' random seeds, \ie{} those on which $A$ errs:
 \[
    B_x \eqdef \setOfSuchThat{ \sigma\in\{0,1\}^\pbits }{ A(x,\sigma) \in\cE }\,.
 \]
 By assumption, $\dabs{B_x} \leq \eta\cdot 2^\pbits$. Now, consider the set $\tilde{B}_x$ of random seeds for which \emph{all} neighbors are bad seeds, that is those seeds for which $A(x,\pi_i(\sigma))$ fails for all $i\in[d]$:
 \[
    \tilde{B}_x \eqdef \setOfSuchThat{ \sigma\in\{0,1\}^\pbits }{ \forall i\in[d]\,,A(x,\pi_i(\sigma)) \in\cE }\,.
 \]
 Since every $\sigma\in \tilde{B}_x$ has $d$ ``bad'' neighbors, we must have $\dabs{e(\tilde{B}_x,B_x)}\geq d\dabs{\tilde{B}_x}$. Applying the Expander Mixing Lemma (\cref{theo:expander:mixing}), we get
 \[
    \frac{\dabs{e(\tilde{B}_x,B_x)}}{dn} \leq \frac{\dabs{B_x}}{n}\cdot \frac{\dabs{\tilde{B}_x}}{n} + \lambda \sqrt{\frac{\dabs{B_x}}{n}\cdot \frac{\dabs{\tilde{B}_x}}{n}}
 \]
 which implies, recalling the above bounds on both $\dabs{\tilde{B}_x}$ and $\dabs{B_x}$, 
$
    \frac{d\dabs{\tilde{B}_x}}{dn} \leq \eta\cdot \frac{\dabs{\tilde{B}_x}}{n} + \lambda \sqrt{\eta\cdot \frac{\dabs{\tilde{B}_x}}{n}}
$. Rearranging,
\[
    \frac{\dabs{\tilde{B}_x}}{n} \leq \lambda^2 \frac{\eta}{(1-\eta)^2}
\]
which is at most $\gamma$ by our choice of $\lambda$. Therefore, for every $x\notin\cX_0$, $\probaDistrOf{\sigma}{x\in \tilde{B}_x} \leq \gamma$, establishing the high-soundness statement.

The bound on $d$, as well as the time efficiency statement, finally follow from the following construction of expanders, due to Bilu and Linial:
\begin{theorem}[{\cite[Theorem 6.12]{HLW06}}]
    For every $d\geq 3$, and every $n\geq 1$, there exists an explicit $d$-regular graph $G$ on $n$ vertices with spectral expansion $\lambda = O( (\log^{3/2} d)/\sqrt{d} )$. Moreover, $G$ can be constructed in time polynomial in $n$ and $d$.
\end{theorem}
\noindent To achieve the desired bound on $\lambda^2$, it therefore suffices to have $d = \tildeO{ \frac{\eta}{(1-\eta)^2\gamma} }$.
\end{proof}
\begin{remark}
  By a probabilistic argument, for all $n,d\geq 1$, and every constant $\delta > 0$, there exist $d$-regular graphs on $n$ vertices with spectral expansion $\lambda \leq (2+\delta)\sqrt{d-1}/d$ (more precisely, almost all $d$-regular graph on $n$ vertices have spectral expansion at most $(2+\delta)\sqrt{d-1}/d$). Therefore, if one does not insist on being able to efficiently construct such a graph, the bound on $d$ in~\cref{theo:deterministic:expansion:restated} can be improved to $d \geq \frac{4.1\eta}{(1-\eta)^2\gamma}$.
\end{remark}

 \section{Omitted proof: Proof of~\cref{theo:semimaxmin}}\label{app:semimaxmin:lb}
In this appendix, we prove of~\cref{theo:semimaxmin}, restated below. 
\begin{theorem}[\cref{theo:semimaxmin}, restated]
  \label{theo:semimaxmin:restated}
Given $\ns\in \N$, $\dst\in (0,1)$, $\pbits\in\N$, for a channel family $\cW$ the
$(\ns,\dst,\pbits)$-semimaxmin chi-square fluctuation is bounded as
\begin{equation*}
\ouchisquaredec(\cW^\ns, \dst, \pbits) 
=
O\left( \frac{\ns^2\dst^4}{\ab^3}\cdot 2^{\pbits} \cdot \max_{W\in \cW}\, \norm{H(W)}_\ast^2\right)\,,
\end{equation*}
whenever
\begin{equation}\label{e:assumption:n:upperbound}
\ns \leq \frac{\ab^{3/2}}{C\dst^2 2^{\pbits/2} \max_{W\in\cW}\, \norm{H(W)}_\ast}\,,
\end{equation}
where $C>0$ is a constant.
\end{theorem} 
\begin{proof}
To obtain the desired bound for semimaxmin chi-square fluctuation, we fix an arbitrary multiset $\cW_\pbits \subseteq\bar{\cW}^\ns$ of size at most $2^\pbits$, and bound the average (over $W^\ns$ in $\cW_\pbits$) decoupled chi-square fluctuation for a
suitable almost $\dst$-perturbation $\cP_\zeta$.  
With this in mind,
suppose we have a random variable $Z=(Z_1,\dots,Z_{\ab/2})$ taking values in
$[-1,1]^{\ab/2}$ and with distribution $\zeta$ such that
\begin{align}
\probaOf{\normone{Z}\geq \frac{\ab}{\beta}}\geq \alpha
\label{e:condition_zeta}
\end{align} 
for some constants $\alpha\geq 1/10$ and $\beta>0$.
For $\dst\in(0,\beta^{-1})$, consider the perturbed family around
$\uniformOn{\ab}$ consisting of elements $\p_z$, $z\in [-1, 1]^{\ab/2}$,
given by
\begin{equation}
\p_z= \frac{1}{\ab} \left(1+\beta\dst z_1, 1-\beta\dst z_1, \ldots, 1+\beta\dst z_t, 1-\beta\dst z_t, \ldots, 1+\beta\dst z_{\ab/2}, 1-\beta\dst z_{\ab/2} \right)\,.
\end{equation}
By our assumption on $Z$ (\cref{e:condition_zeta}), $\p_Z$ then satisfies 
$\totalvardist{\p_Z}{\uniformOn{\ab}} = 
\frac{\beta\dst}{\ab}\normone{Z} \geq \dst$
with probability at least $\alpha$. 
Consider any $W^\ns\in\cW_\pbits^\ns$. 
From the same steps as in~\cite[Theorem IV.14]{ACT:18:IT1}, we get
\begin{equation}
\chisquaredec\left(W^\ns \mid \cP_\zeta\right)
= \ln \bE{ZZ^\prime}{\exp\left( \frac{\beta^2\ns\dst^2}{\ab}\cdot
Z^\transp\bar{H}(W^\ns) Z'\right)},
\end{equation}
where $Z,Z'$ are independent random variables with common distribution
$\zeta$, $\bar{H}(W^\ns) \eqdef \frac{1}{\ns}\sum_{j=1}^\ns H(W_j)$, and $H(W_j)$ is defined as in~\cref{eq:def:barH}. Now, to bound the semimaxmin chi-square fluctuation, we must handle $\shortexpect_{W^\ns}[\chisquaredec\left(W^\ns \mid \cP_\zeta\right)]$, for $W^\ns$ drawn uniformly at random from $\cW_\pbits$. We thus define the new ``aggregate'' matrix
\[
    \tilde{H}(\cW_{\pbits}) \eqdef \sum_{W^\ns\in\cW_{\pbits}} \overline{H}(W^\ns)
\]
to which we apply a construction of Acharya, Canonne, and Tyagi, whose properties we summarize below.
\begin{lemma}[{Implicit in the proof of~\cite[Theorem IV.18]{ACT:18:IT1}}]
    Let $A\in\R^{(\ab/2)\times(\ab/2)}$ be a p.s.d. matrix. Then, there exists a matrix $V\in\R^{(\ab/2)\times(\ab/4)}$ such that the following holds.
    \begin{enumerate}[(i)]
      \item Each row vector of $V$ has $\lp[2]$ norm at most $1$, and $V$ has Frobenius norm $\norm{V}_F \geq \sqrt{\ab}/2$.
      \item\label{lemma:construction:V:item2} Let $Y=(Y_1\ldots Y_{\ab/4})$ be a vector of i.i.d. Rademacher random variables. Then,\footnote{We note that this second item is a consequence of the first, along with Khintchine's inequality and an anticoncentration argument; see~\cite[Claim IV.21]{ACT:18:IT1}. For clarity, we nonetheless explicitly state both here.}
      \[
          \probaOf{\normone{VY}\geq \frac{\ab}{12\sqrt{2}}}\geq \frac{1}{9}.
      \]
      \item\label{lemma:construction:V:item3} We have $\norm{V^\transp AV}_F^2 \leq \frac{4}{\ab}\norm{A}_\ast^2$.
    \end{enumerate}
\end{lemma}
We invoke this lemma on $\tilde{H}(\cW_{\pbits})$, and denote by $V\in\R^{(\ab/2)\times(\ab/4)}$ the resulting matrix. Letting $\zeta$ be the distribution of the random variable $Z\eqdef VY$, where $Y$ is a vector of $\ab/4$ i.i.d. Rademacher random variables, item~\ref{lemma:construction:V:item2} implies that $\zeta$ satisfies the condition from~\cref{e:condition_zeta}, for $\alpha \eqdef 1/9$ and $c\eqdef 1/(12\sqrt{2})$. Moreover, since all $\overline{H}(W^\ns)$ (and therefore all $V^\transp \overline{H}(W^\ns) V$'s) are symmetric p.s.d. matrices, we have\footnote{This follows from the fact that $\Tr AB \geq 0$ for two p.s.d. matrices $A,B$; and thus $\norm{A+B}_F^2 = \Tr[(A+B)^2] = \Tr[A^2]+\Tr[B^2] + 2\Tr[AB] \geq \norm{A}_F^2 + \norm{B}_F^2$.}
\[
    \norm{V^\transp \sum_{W^\ns\in\cW_{\pbits}} \overline{H}(W^\ns) V }_F^2 \geq \sum_{W^\ns\in\cW_{\pbits}}  \norm{V^\transp \overline{H}(W^\ns) V }_F^2\,,
\]
or, equivalently,
$
    2^{-\pbits} \norm{V^\transp \tilde{H}(\cW_{\pbits}) V }_F^2 \geq \bE{W^\ns}{\norm{V^\transp \overline{H}(W^\ns) V}_F^2}
$.
However, by item~\ref{lemma:construction:V:item3}, $\norm{V^\transp \tilde{H}(\cW_{\pbits}) V}_F^2 \leq (4/\ab)\norm{\tilde{H}(\cW_{\pbits})}_\ast^2$. Since, by the triangle inequality,  we further have 
\[
  \norm{\tilde{H}(\cW_{\pbits})}_\ast \leq 2^\pbits \max_{W^\ns\in\cW_\pbits} \norm{\overline{H}(W^\ns)}_\ast \leq 2^\pbits \max_{W\in \cW} \norm{H(W)}_\ast
\]
we obtain
\begin{equation}\label{eq:semimaxmin:crucial:bound:V}
    \bE{W^\ns}{\norm{V^\transp \overline{H}(W^\ns) V}_F^2} \leq \frac{4\cdot 2^\pbits}{\ab}\max_{W\in \cW} \norm{H(W)}_\ast^2\,.
\end{equation}
We can now bound $\shortexpect_{W^\ns}[\chisquaredec\left(W^\ns \mid \cP_\zeta\right)]$. Let $c>0$ be the constant from the statement of~\cref{l:ic_testing_fluctuation_bound_semiprivate}. 
Setting $\lambda\eqdef (\beta^2\ns\dst^2)/\ab$ and recalling our assumption (\cref{e:assumption:n:upperbound}) on $\ns$, we have
\begin{align*}
1 &\geq \frac{16\beta^2\ns \dst^2 2^{\pbits/2} \cdot \max_{W\in \cW} \norm{H(W)}_\ast}{c\ab^{3/2}}
\geq \frac{8\lambda}{c} \sqrt{ \bE{W^\ns}{\norm{V^\transp \overline{H}(W^\ns) V}_F^2}  }
\geq \frac{8\lambda}{c} \bE{W^\ns}{\norm{V^\transp \overline{H}(W^\ns) V}_F } \\
&\geq \frac{8\lambda}{c} \bE{W^\ns}{\rho(V^\transp \bar{H}(W^\ns) V)}\,,
\end{align*}
where the second inequality is~\cref{eq:semimaxmin:crucial:bound:V} and $\rho(A)$ denotes the spectral norm of matrix $A$. By Markov's inequality, we have that 
\[
    \probaDistrOf{W^\ns}{ \rho(V^\transp \bar{H}(W^\ns) V) > \frac{1}{4\lambda} } \leq \frac{c}{2}\,.
\]
Let $\goodset\subseteq \cW_\pbits$ (for ``good'') be the multiset of $W^\ns$'s such that $\rho(V^\transp \bar{H}(W^\ns) V) \leq 1/(4\lambda)$, which by the above has size at least $(1-c/2)\cdot 2^\pbits$. Upon reorganizing, for any $W^\ns\in \goodset$ we have
\[
\lambda^2/(1-4\lambda^2 \rho(V^\transp \bar{H}(W^\ns) V)^2) \leq 4\lambda^2/3\,.\]
We can then apply the lemma below on the MGF of a Rademacher chaos to i.i.d. Rademacher random variables $Y$ and the symmetric matrix $V^\transp \bar{H}(W^\ns) V\in\R^{\ab/4\times\ab/4}$:
\begin{lemma}[{\cite[Claim IV.17]{ACT:18:IT1}}]\label{claim:mgf:bound:chaos}
For random vectors $\theta, \theta'\in\{-1, 1\}^{\ab/2}$ with each
$\theta_i$ and $\theta'_i$ distributed uniformly over $\{-1, 1\}$,
independent of each other and independent for different $i$'s. Then, for a positive semi-definite matrix $H$,
\[
\ln \bE{\theta \theta'}{ e^{\lambda\theta^\transp
H\theta'}} \leq \frac{\lambda^2}{2}\cdot\frac{\norm{H}_F^2}{1-4\lambda^2\rho(H)^2} \,, \quad \forall\,
0\leq \lambda <\frac 1{2\rho(H)},
\]
where $\norm{\cdot}_F$ denotes the Frobenius norm and $\rho(\cdot)$
the spectral radius.
\end{lemma}
\noindent This gives, for any $W^\ns\in\goodset$,
\begin{equation*}\label{eq:bound:frobenius}
    \bE{ZZ^\prime}{\exp\mleft( \tfrac{\beta^2\ns\dst^2}{\ab}\cdot Z^\transp\bar{H}(W^\ns)Z'\mright)} = \shortexpect_{YY'}[ e^{\frac{\beta^2\ns\dst^2}{\ab}Y^\transp
    V^\transp \bar{H}(W^\ns) VY'} ] \leq
    e^{\frac{2\beta^4\ns^2\dst^4}{3\ab^2}\norm{V^\transp \bar{H}(W^\ns) V}_F^2}\,.
\end{equation*}
From there, by concavity and using Jensen's inequality, we obtain
\begin{align*}
    \shortexpect_{W^\ns}[1\land \chisquaredec\left(W^\ns \mid \cP_\zeta\right)]
    &\leq \shortexpect_{W^\ns}[\chisquaredec\left(W^\ns \mid \cP_\zeta\right)\indicSet{\goodset}(W^\ns) + \indicSet{\goodset^c}(W^\ns)] \\
    &\leq \shortexpect_{W^\ns}[\chisquaredec\left(W^\ns \mid \cP_\zeta\right)\indicSet{\goodset}(W^\ns)] + \frac{c}{2}\\
    &\leq \shortexpect_{W^\ns}[\ln( e^{\frac{2\beta^4\ns^2\dst^4}{3\ab^2}\norm{V^\transp \bar{H}(W^\ns) V}_F^2} )]+ \frac{c}{2} \\
    &\leq \ln( e^{\frac{2\beta^4\ns^2\dst^4}{3\ab^2}\shortexpect_{W^\ns}[\norm{V^\transp \bar{H}(W^\ns) V}_F^2]} ) + \frac{c}{2} \,.
\end{align*}
The above, along with~\cref{eq:semimaxmin:crucial:bound:V}, then finally yields
\begin{equation*}
    \shortexpect_{W^\ns}[1\land \chisquaredec\left(W^\ns \mid \cP_\zeta\right)]
    \leq \frac{8\beta^4\ns^2\dst^4 2^\pbits}{3\ab^3}\max_{W\in \cW} \norm{H(W)}_\ast^2+ \frac{c}{2}\,,
\end{equation*}
which, invoking~\cref{l:ic_testing_fluctuation_bound_semiprivate}, completes the proof.
\end{proof}

\end{document}